\documentclass[twocolumn,showpacs,preprintnumbers,amsmath,amssymb,prb,superscriptaddress]{revtex4}
\usepackage{times}
\usepackage{amsfonts}
\usepackage{mathrsfs}
\usepackage{graphicx}
\usepackage{dcolumn}
\usepackage{bm}
\usepackage{color}

\usepackage[colorlinks,bookmarks=false,citecolor=blue,linkcolor=red,urlcolor=blue]{hyperref}
\usepackage{multirow}
\definecolor{ZYcolor}{rgb}{0.1,0.5,0.4}

\newcommand{\Rom}[1]{ \uppercase\expandafter{\romannumeral#1}}

\def\be{\begin{equation}}       \def\ee{\end{equation}}
\def\bea{\begin{eqnarray}}      \def\eea{\end{eqnarray}}

\begin{document}


\title{ Spin-triplet Superconductivity in Nonsymmorphic crystals }

\author{Shengshan Qin}\email{qinshengshan@ucas.ac.cn}
\affiliation{Kavli Institute for Theoretical Sciences and CAS Center for Excellence in Topological Quantum Computation, University of Chinese Academy of Sciences, Beijing 100190, China}

\author{Chen Fang}
\affiliation{Beijing National Research Center for Condensed Matter Physics,
and Institute of Physics, Chinese Academy of Sciences, Beijing 100190, China}
\affiliation{Kavli Institute for Theoretical Sciences and CAS Center for Excellence in Topological Quantum Computation, University of Chinese Academy of Sciences, Beijing 100190, China}

\author{Fu-chun Zhang}
\affiliation{Kavli Institute for Theoretical Sciences and CAS Center for Excellence in Topological Quantum Computation, University of Chinese Academy of Sciences, Beijing 100190, China}
\affiliation{Collaborative Innovation Center of Advanced Microstructures, Nanjing University, Nanjing 210093, China}

\author{Jiangping Hu}\email{jphu@iphy.ac.cn}
\affiliation{Beijing National Research Center for Condensed Matter Physics,
and Institute of Physics, Chinese Academy of Sciences, Beijing 100190, China}
\affiliation{Kavli Institute for Theoretical Sciences and CAS Center for Excellence in Topological Quantum Computation, University of Chinese Academy of Sciences, Beijing 100190, China}
\affiliation{South Bay Interdisciplinary Science Center, Dongguan, Guangdong Province, China}

\date{\today}

\begin{abstract}

Spin-triplet superconductivity is known to be a rare quantum phenomenon. Here we show that  nonsymmorphic crystalline symmetries can dramatically assist spin-triplet superconductivity in the presence of spin-orbit coupling. Even with a weak spin-orbit coupling, the spin-triplet pairing can be the leading pairing instability in a lattice with a nonsymmorphic symmetry. The underlining mechanism is the \textit{spin-sublattice-momentum lock}  on electronic bands that are protected by the nonsymmorphic symmetry. We use the nonsymmorphic space group $P4/nmm$ to demonstrate these results and discuss related experimental observables.  Our work paves a new way in searching for spin-triplet superconductivity.
\end{abstract}

\maketitle
\textit{Introduction.}
The spin-triplet superconductors, which are the superconducting analogy of the $^3$He superfluid\cite{Book_He3}, have been long-pursued. They have been proposed to be natural candidates for the topological superconductors\cite{RevModPhys.82.3045, RevModPhys.83.1057, RevModPhys.88.035005, Alicea_2012, Kitaev_2001, hao2019topological, wu2020pursuit}, hosting the Majorana modes which are expected to play an essential role in the fault-tolerant quantum computations\cite{RevModPhys.80.1083, PhysRevX.5.041038, lian2018topological}. In the past decades, great efforts have been made in pursuing the spin-triplet superconductors\cite{RevModPhys.63.239, RevModPhys.75.657, noncentro_SC}. Theoretically, various mechanisms have been proposed in favor of the spin-triplet superconductivity. For instance, the spin-triplet superconductivity may arise at ultra low temperature through the Kohn-Luttinger mechanism\cite{PhysRevLett.15.524}; and it can also be induced from the ferromagnetic spin fluctuations or the ferromagnetic exchange coupling\cite{RevModPhys.75.657}. Experimentally, Sr$_2$RuO$_4$ has been suggested to be a promising candidate for the spin-triplet superconductors\cite{SrRuO, SrRuO_sigrist}. However, recent experiments raise doubts on this issue\cite{SrRuO_doubt}. In many heavy fermion system such as UPt$_3$\cite{RevModPhys.74.235}, UTe$_2$\cite{UTe2}, and the recently synthesized K$_2$Cr$_3$As$_3$\cite{PhysRevX.5.011013, KCrAs_zheng}, signatures for the spin-triplet superconductivity have been observed.

During the past decades, the spin-orbit coupled systems have attracted more and more research attentions. The spin-orbit coupling (SOC) has been revealed to play an important role in various exotic condensed matter systems such as the topological materials\cite{RevModPhys.88.021004, JPSJ_TI, annual_TCI, RevModPhys.90.015001, annual_WS}. Recent studies suggest that the SOC can help the spin-triplet superconductivity. For instance, it has been predicted the spin-triplet superconductivity may exist in doped superconducting topological insulators\cite{PhysRevLett.105.097001, PhysRevB.90.100509, PhysRevB.90.184516, PhysRevB.94.180504, PhysRevX.8.041026} and semimetals\cite{PhysRevLett.115.187001, PhysRevB.94.014510}. Especially, in the doped topological insulator Bi$_2$Se$_3$ nematic superconductivity has been confirmed experimentally\cite{TI_nemetic1, TI_nemetic2, TI_nemetic3, TI_nemetic4, PhysRevX.7.011009, PhysRevX.8.041024}, indicating possible odd-parity spin-triplet superconductivity in the system\cite{PhysRevLett.105.097001, PhysRevB.90.100509}. Besides the topological materials, in the two-dimensional (2D) electron gas formed at the interface between LaAlO$_3$ and SrTiO$_3$\cite{interface1, PhysRevLett.104.126803}, the spin-triplet superconductivity is also proposed based on large Rashba SOC\cite{PhysRevLett.108.147003, PhysRevB.80.140509}. However, all these studies rely on a strong SOC, in which  pairing forces may  be significantly weakened by the SOC as well.

In this Letter, we show that the spin-triplet superconductivity can be stabilized by nonsymmorphic symmetries in the spin-orbit coupled systems. Even with a  weak SOC, the spin-triplet pairing can be the leading pairing instability in a lattice with a nonsymmorphic symmetry. We specify our study with the nonsymmorphic space group $P4/nmm$ ($\#.129$).  Due to the nonsymmorphic symmetries, the sublattice degree always exists in the system. In the presence of the SOC, the sublattice degree intertwines with the spin degree. Correspondingly, the spin, sublattice and momentum are locked with each other, forming a \textit{spin-sublattice-momentum lock} texture on the normal-state energy bands. The spin-triplet pairing state is always favored due to spin-sublattice-momentum lock when there is a pairing force between the two sublattices.

\begin{figure}[!htbp]
	\centering
	\includegraphics[width=0.98\linewidth]{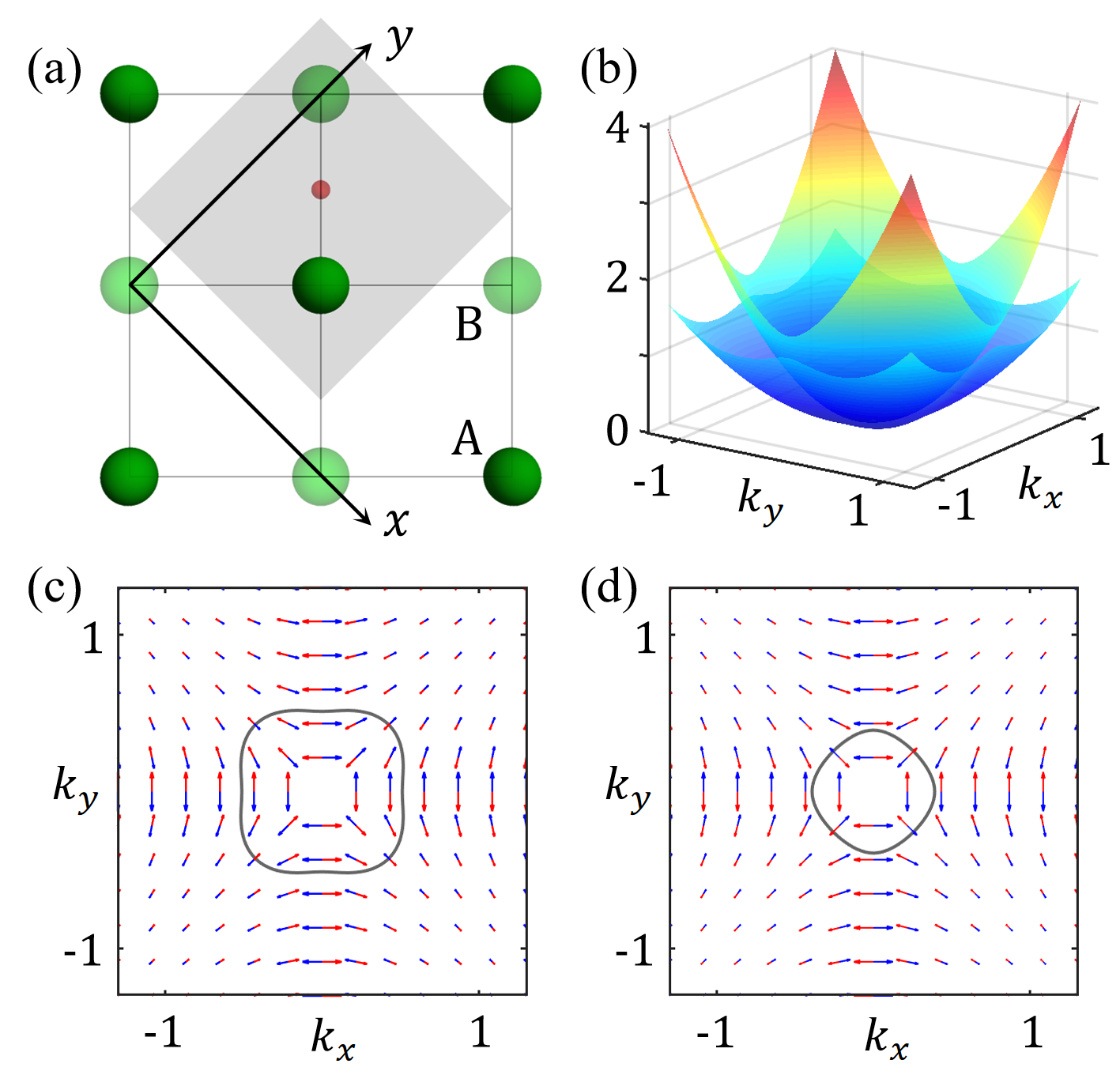}
	\caption{\label{fig1} (color online) (a) A sketched quasi-2D lattice structure respecting the $P4/nmm$ space group:  A and B indicate the sublattices related by the nonsymmorphic symmetries, the shadow region indicates the unit cell, and the red point is the inversion center located at the bond center between two nearest neighbouring sites. (b) The band structure near the M point, plotted from the Hamiltonian in Eq.\eqref{normal_Hamiltonian_kp_main} with parameters $\{ t, t^\prime, \lambda \} = \{ 1.0, 0.8, 0.12 \}$. (c) and (d) show the spin polarizations on the lower and upper energy bands in (b) respectively: the spin polarization contributed by the A (B) sublattice is labeled by the red (blue) arrowed line, with the length of the line indicating the strength of the polarization. The gray lines in (c)(d) show the Fermi surfaces for chemical potential $\mu = 0.2$.}
\end{figure}

We first briefly review the space group $\mathcal{G} = P4/nmm$, which is nonsymmorphic. There are 16 symmetry operations in its quotient group $\mathcal{G}/T$ with $T$ being the translation group. More specially, $\mathcal{G}/T$ can be written in the following concise direct product form\cite{PhysRevX.3.031004}
\begin{align}\label{quotientG_main}
\begin{split}
\mathcal{G}/T &= D_{2d} \otimes Z_2,
\end{split}
\end{align}
in a sense that symmetry operations are equivalent if they differ by a lattice translation. We specify the symmetry group with a quasi-2D lattice shown in Fig.\ref{fig1}(a), which is similar to the structure of the monolayer FeSe. As shown in the lattice, the fixed point of point group $D_{2d}$ in Eq.\eqref{quotientG_main} is at the lattice sites, and $Z_2$ is a two-element group including the inversion symmetry which is defined at the bond center between two nearest lattice sites. According to Eq.\eqref{quotientG_main}, we can choose the generators of the quotient group $\mathcal{G}/T$ as the inversion symmetry $\{ I | {\bf \tau}_0 \}$, the mirror symmetry $\{ M_y | {\bf 0} \}$ and the rotoinversion symmetry $\{ S_{4z} | {\bf 0} \}$\cite{footnote0}, where the symmetry operators have been expressed in the form of the Seitz operators and ${\bf \tau}_0 = {\bf a}_1/2 + {\bf a}_2/2$ with ${\bf a}_1$ and ${\bf a}_2$ being the primitive lattice translations along the $x$ and $y$ directions in Fig.\ref{fig1}(a).

\textit{Low-energy theory near $(\pi, \pi)$.}
A standard group theory analysis shows that the space group $P4/nmm$ merely has one single 4D irreducible representation at the Brillouin zone corner $(\pi, \pi)$, i.e. the M point, in the spinful condition\cite{TSC_band_degeneracy}. The above conclusion straightforwardly leads to three important implications. (i) For systems respecting the space group $P4/nmm$, in the presence of SOC all the energy bands are fourfold degenerate at the M point. (ii) All the fourfold degenerate bands respect the same low-energy effective model. (iii) One can use arbitrary orbital to construct the low-energy effective theory near M, and for simplicity we consider one $s$ orbital at each lattice site in Fig.\ref{fig1}(a) in the following.

With the above preparation, we can construct the low-energy effective theory near M. As all the symmetry operations in $\mathcal{G}/T$ preserve at $(\pi, \pi)$, we need to derive the matrix form of the symmetry generators of $\mathcal{G}/T$. By a careful analysis, we obtain the matrix form of the symmetry operators as, $\mathcal{I} = s_0 \sigma_1$, $\mathcal{M}_y = i s_2 \sigma_3$ and $\mathcal{S}_{4z} = e^{ is_3\pi/4 } \sigma_3$, where $\mathcal{I}$, $\mathcal{M}_y$ and $\mathcal{S}_{4z}$ stand for $\{ I | {\bf \tau_0} \}$, $\{ M_y | {\bf 0} \}$ and $\{ S_{4z} | {\bf 0} \}$ respectively (details in SM). In the matrix form, $s_i$ and $\sigma_i$ $(i = 1, 2, 3)$ are the Pauli matrices for the spin and the two sublattices respectively, and $s_0$ and $\sigma_0$ the corresponding identity matrices. The above matrices are actually a set of irreducible representation matrices for space group $P4/nmm$ at M. Besides the crystalline symmetries, the time reversal symmetry is $\mathcal{T} = i s_2 \sigma_0 K$ with $K$ the complex conjugation operation.

The low-energy effective theory near M for group $P4/nmm$ is generally depicted by the sixteen $\Gamma = s_i \sigma_j$ matrices. In deriving the effective model, it is convenient to first constrain the system by the time reversal symmetry and the inversion symmetry, and then consider the constraints of other crystalline symmetries. After some algebra, we classify the symmetry allowed $\Gamma$ matrices along with the ${\bf k}$-dependent functions, and obtain the low-energy effective Hamiltonian as follows\cite{footnote1} (details in SM)
\begin{eqnarray}\label{normal_Hamiltonian_kp_main}
\mathcal{H}_{\text eff}({\bf k}) &=& m({\bf k}) s_0 \sigma_0 + \lambda k_x s_2 \sigma_3 + \lambda k_y s_1 \sigma_3  \nonumber \\
& & + t^\prime k_x k_y s_0 \sigma_1,
\end{eqnarray}
where $m({\bf k}) = t (k_x^2 + k_y^2)$. Notice that $k_{x/y}$ is defined according to the M point here. To have a more intuitive impression on the effective theory in Eq.\eqref{normal_Hamiltonian_kp_main}, one can understand the parameters in the lattice shown in Fig.\ref{fig1}(a). Specifically, $t$ ($t^\prime$) describes the hopping between the intrasublattice (intersublattice) nearest neighbours, and $\lambda$ is the inversion-symmetric Rashba SOC arising from the mismatch between the lattice sites and the inversion center\cite{PhysRevX.12.011030}, i.e. the local inversion-symmetry breaking\cite{local_inversion}. We show the band structures calculated from the effective Hamiltonian in Eq.\eqref{normal_Hamiltonian_kp_main} in Fig.\ref{fig1}(b). Due to the presence of both the time reversal and inversion symmetries, all the energy bands are twofold degenerate.

\textit{Spin-sublattice-momentum lock.}
In centrosymmetric systems, the local inversion-symmetry breaking can intertwine the different degrees of freedom\cite{zhang2014hidden, wu2017direct, Zhang_2020}. Here, for systems respecting the space group $P4/nmm$, symmetries enforce the spin degree locked to the sublattice degree on the energy bands and the sublattice-distinguished spin is nearly fully polarized for small Fermi surfaces near $( \pi, \pi)$. Before the detailed calculations, we first consider the symmetry constraints. As shown in Fig.\ref{fig1}(a), the inversion symmetry exchanges the two sublattices, while the time reversal symmetry does not change the spacial position. On the other hand, the inversion symmetry preserves the spin but the time reversal symmetry flips the spin. Therefore, considering the combination of the time reversal symmetry and inversion symmetry, at each ${\bf k}$ point the spin polarizations from the two sublattices are always opposite. Moreover, due to the mirror symmetry $\{ M_{x/y} | {\bf 0} \}$, the spin is polarized perpendicular to the mirror plane along $k_{x/y} = 0$, i.e. the Brillouin zone boundary.

Based on the low-energy effective theory in Eq.\eqref{normal_Hamiltonian_kp_main}, the spin polarizations contributed by the different sublattices, i.e. $\langle {\bf s}_A \rangle$ and $\langle {\bf s}_B \rangle$, can be calculated analytically (details in SM). A direct calculation shows that $\langle {\bf s}_B \rangle = -\langle {\bf s}_A \rangle = ( \sin\theta, \cos\theta, 0 ) / (\sqrt{ 1 + t^{\prime 2} k^2 \sin^2 2\theta / 4 \lambda^2 } )$ at point ${\bf k}$, with ${\bf k}$ written as $(k_x, k_y) = (k\cos\theta, k\sin\theta)$. We sketch the results in Fig.\ref{fig1}(c)(d). As shown, both $\langle {\bf s}_A \rangle$ and $\langle {\bf s}_B \rangle$ lie in the $xy$ plane and wind around the M point anticlockwise. Moreover, the spin polarization reaches its minimum along $k_x = k_y$, and is nearly fully polarized on the bands near the Brillouin zone boundary. It is worth pointing out that, the fully polarized spin along the Brillouin zone boundary satisfying $\langle {\bf s}_A \rangle = - \langle {\bf s}_B \rangle$ is consistent with the matrix form of the mirror symmetries at M, i.e. $\mathcal{M}_{x} = i s_{1} \sigma_3$ and $\mathcal{M}_{y} = i s_{2} \sigma_3$\cite{footnote2}.

\textit{Superconductivity.}
For the effective theory in Eq.\eqref{normal_Hamiltonian_kp_main}, we consider the possible superconductivity induced by the phenomenological density-density interactions
\begin{eqnarray}\label{interaction_main}
\mathcal{H}_{int} = \int d{\bf q} [ U \sum_{i=1}^2 n_i({\bf q}) n_i(-{\bf q}) + 2V n_1({\bf q}) n_2(-{\bf q}) ],
\end{eqnarray}
where $n_1({\bf q}) = \sum_{\kappa=\uparrow,\downarrow} \frac{1}{ \sqrt{N} } \int d{\bf k} c^\dagger_{{\bf k},\kappa} c_{{\bf k + q},\kappa}$ and $n_2({\bf q}) = \sum_{\kappa=\uparrow,\downarrow} \frac{1}{ \sqrt{N} } \int d{\bf k} d^\dagger_{{\bf k},\kappa} d_{{\bf k + q},\kappa}$ are the density operators for the A and B sublattices in Fig.\ref{fig1}(a) respectively, and $U$ and $V$ are the intrasublattice and intersublattice interactions respectively. In Eq.\eqref{interaction_main}, we focus on the momentum-independent interactions in the weak-coupling condition. Obviously, the negative $U$ ($V$) correspond to the attractive interaction. Actually, the phenomenological interactions in Eq.\eqref{interaction_main} arise from the short-range density-density interactions in the real space. Specifically, $U$ is the onsite interaction, and $V$ is the leading-order term, i.e. the momentum-independent part, in the intersublattice interaction between nearest neighbours (details in SM).

From the interactions in Eq.\eqref{interaction_main}, in the mean-field level only the momentum-independent superconducting orders are expected\cite{footnote3}. Due to the fermionic statistics of electrons, the pairing orders are required to satisfy $\hat{\Delta} (-{\bf k}) \cdot is_2\sigma_0 = - ( \hat{\Delta} ({\bf k}) \cdot is_2\sigma_0 )^T$, where the pairing term is $\psi^\dag({\bf k}) \hat{\Delta} ({\bf k}) \cdot is_2\sigma_0 \psi^\dag(-{\bf k})$ in the basis $\psi^\dag({\bf k}) = (c_{{\bf k},\uparrow}^\dag, c_{{\bf k},\downarrow}^\dag, d_{{\bf k},\uparrow}^\dag, d_{{\bf k},\downarrow}^\dag)$. The pairing orders can be further classified in accordance with the symmetry group of the system, and we classify the momentum-independent pairing orders and present the results in Table.\ref{classification_pair_main}. As shown, the pairing orders belong to five different pairing symmetries in the $A_{1g}$, $B_{2g}$, $A_{2u}$, $B_{2u}$ and $E_u$ representations of the $D_{4h}$ group, with the $A$ and $B$ representations being 1D and the $E$ representation 2D. To show the meaning of the pairing orders clear, we list the explicit form of the superconducting pairing as follows
\begin{eqnarray}\label{pair_explicit}
\hat{\Delta}_{ A_{1g} } &:&  c^\dagger_{ {\bf k}, \uparrow } c^\dagger_{ -{\bf k}, \downarrow } + d^\dagger_{ {\bf k}, \uparrow } d^\dagger_{ -{\bf k}, \downarrow },  \\
\hat{\Delta}_{ B_{2g} } &:&  c^\dagger_{ {\bf k}, \uparrow } d^\dagger_{ -{\bf k}, \downarrow } + d^\dagger_{ {\bf k}, \uparrow } c^\dagger_{ -{\bf k}, \downarrow },  \nonumber \\
\hat{\Delta}_{ A_{2u} } &:&  -ic^\dagger_{ {\bf k}, \uparrow } d^\dagger_{ -{\bf k}, \downarrow } + id^\dagger_{ {\bf k}, \uparrow } c^\dagger_{ -{\bf k}, \downarrow },  \nonumber \\
\hat{\Delta}_{ B_{2u} } &:&  c^\dagger_{ {\bf k}, \uparrow } c^\dagger_{ -{\bf k}, \downarrow } - d^\dagger_{ {\bf k}, \uparrow } d^\dagger_{ -{\bf k}, \downarrow },  \nonumber \\
\hat{\Delta}_{ E_u }    &:&  ( ic^\dagger_{ {\bf k}, \uparrow } d^\dagger_{ -{\bf k}, \uparrow } - ic^\dagger_{ {\bf k}, \downarrow } d^\dagger_{ -{\bf k}, \downarrow }, c^\dagger_{ {\bf k}, \uparrow } d^\dagger_{ -{\bf k}, \uparrow } + c^\dagger_{ {\bf k}, \downarrow } d^\dagger_{ -{\bf k}, \downarrow } ). \nonumber
\end{eqnarray}
As mentioned, the inversion symmetry in space group $P4/nmm$ exchanges the two sublattices. $\hat{\Delta}_{ A_{1g} }$ and $\hat{\Delta}_{ B_{2g} }$ are spin-singlet pairings with even parity, and $\hat{\Delta}_{ A_{1g} }$ occurs in the same sublattice while $\hat{\Delta}_{ B_{2g} }$ is between the different sublattices. $\hat{\Delta}_{ B_{2u} }$ is the intrasublattice spin-singlet pairing with odd parity, whereas $\hat{\Delta}_{ A_{2u} }$ and $\hat{\Delta}_{ E_u }$ are the odd-parity spin-triplet pairings between the different sublattices.

\begin{table}[]
\caption{\label{classification_pair_main}  Classification of the possible momentum-independent pairing potentials corresponding to the interactions in Eq.\eqref{interaction_main}, according to the irreducible representations of the $D_{4h}$ point group. Here, the pairing potentials are in the form $\psi^\dag({\bf k}) \hat{\Delta} \cdot is_2\sigma_0 \psi^\dag(-{\bf k})$, with the basis being $\psi^\dag({\bf k}) = (c_{{\bf k},\uparrow}^\dag, c_{{\bf k},\downarrow}^\dag, d_{{\bf k},\uparrow}^\dag, d_{{\bf k},\downarrow}^\dag)$.  }
\begin{tabular}{ p{1.0cm}<{\centering} p{1.2cm}<{\centering} p{1.2cm}<{\centering} p{1.2cm}<{\centering} p{1.2cm}<{\centering} p{2.0cm}<{\centering} } \hline\hline
          &   $E$   &   $\mathcal{S}_{4z}$   &   $\mathcal{I}$   &   $\mathcal{M}_y$     &   $\hat{\Delta}$   \\ \hline
$A_{1g}$  &  1  &  1   &  1   &  1     &  $s_0 \sigma_0$  \\
$B_{2g}$  &  1  &  -1  &  1   &  -1    &  $s_0 \sigma_1$  \\
$A_{2u}$  &  1  &  -1  &  -1  &  1     &  $s_3 \sigma_2$  \\
$B_{2u}$  &  1  &  1   &  -1  &  1     &  $s_0 \sigma_3$  \\
$E_u$     &  2  &  0   &  -2  &  0     &  $( s_1 \sigma_2, s_2 \sigma_2)$  \\ \hline\hline
\end{tabular}
\end{table}

To find out the superconducting ground state, we solve the following linearized gap equations (details in SM)
\begin{eqnarray}\label{gap_equation_main}
\hat{\Delta}_{ A_{1g}, B_{2u} } &:&  -U \chi_{ A_{1g}, B_{2u} }(T_c) = 1, \\
\hat{\Delta}_{ B_{2g}, A_{2u}, E_u } &:&  -V \chi_{ B_{2g}, A_{2u}, E_u }(T_c) = 1,  \nonumber
\end{eqnarray}
where we have used the fact that $\hat{\Delta}_{ A_{1g} }$ and $\hat{\Delta}_{ B_{2u} }$ can only result from the intrasublattice interaction $U$, and $\hat{\Delta}_{ B_{2g} }$, $\hat{\Delta}_{ A_{2u} }$ and $\hat{\Delta}_{ E_u }$ only arise from the intersublattice interaction $V$. In Eq.\eqref{gap_equation_main}, $\chi$ is the finite-temperature superconducting susceptibility for each irreducible representation pairing channel in Table.\ref{classification_pair_main}, which can be calculated as
\begin{eqnarray}\label{susceptibility_sc_main}
\chi(T_c) = \mathcal{F}(T_c) \sum_s \int d\theta D(\theta) \sum_{ s^\prime=s,\bar{s} }  | \langle u_{s, {\bf k}} | \hat{\Delta} | u_{{s^\prime}, {\bf k}} \rangle |^2.
\end{eqnarray}
In the above equation, $| u_{ s, {\bf k}} \rangle$ is the wavefunction for the state on the Fermi surface contributed by band $s$, and $| u_{\bar{s}, {\bf k}} \rangle = \mathcal{IT} | u_{ s, {\bf k}} \rangle$ is the state degenerate with $| u_{ s, {\bf k}} \rangle$ due to the presence of both the inversion symmetry $\mathcal{I}$ and the time reversal symmetry $\mathcal{T}$. $\mathcal{F}(T_c) = \frac{1}{2N}\int^{\omega_0}_{-\omega_0}  \frac{1}{2\xi} \tanh \frac{\beta\xi}{2}  d\xi$ is a temperature-dependent constant with $\beta = 1/k_BT_c$ and $\omega_0$ the energy cutoff near the Fermi energy. $D(\theta) = 2dk^\prime / d\xi_{s, {\bf k^\prime} }$ is the density of states on the Fermi surface. By solving Eq.\eqref{gap_equation_main}, we can get the superconducting transition temperature for each pairing channel, and the state with the highest $T_c$ is the ground state.

\begin{figure}[!htbp]
	\centering
	\includegraphics[width=0.75\linewidth]{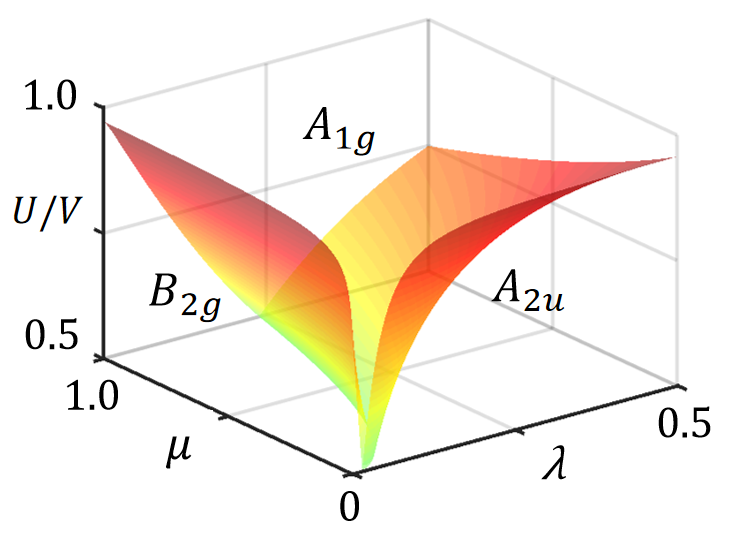}
	\caption{\label{fig2} (color online) Superconducting phase diagram versus the chemical potential $\mu$, the SOC  $\lambda$, and the interaction $U/V$, assuming $U$ and $V$ both attractive. The colored surface in the figure is the phase boundary between the different superconducting ground states. In the calculation, the other parameters are $\{ t, t^\prime \} = \{ 1.0, 0.8 \}$. Here, only the condition for $2|U| > |V|$ is shown. For the following two conditions, (i) $2|U| < |V|$ and (ii) $U>0$ and $V < 0$, only the $B_{2g}$ and $A_{2u}$ states can appear in the phase diagram, with their phase boundary always the same with that at $2U = V$. }
\end{figure}

According to Eqs.\eqref{gap_equation_main}\eqref{susceptibility_sc_main}, the superconducting instability can merely arise from the attractive interactions. Moreover, a direct calculation shows that the superconducting susceptibilities always satisfies $\chi_{ B_{2u} } < \chi_{ A_{1g} }$ and $\chi_{ A_{2u} } = 2 \chi_{ E_u }$ (details in SM), meaning that the $B_{2u}$ and $E_u$ pairing states can never be the ground states. Consequently, in the condition with $U < 0$ and $V > 0$, i.e. the intrasublattice attractive and intersublattice repulsive interactions, the ground state is always the $A_{1g}$ state; and in the condition with $U > 0$ and $V < 0$, the $B_{2g}$ and $A_{2u}$ states can be the superconducting ground states. If both of the interactions are attractive, the $A_{1g}$, $B_{2g}$ and $A_{2u}$ pairing states can appear in different regions in the parameter space, and the corresponding phase diagram is presented in Fig.\ref{fig2}. In the phase diagram, we merely show the condition for $U/V > 0.5$. Whereas, the phase diagram for $U/V < 0.5$ is independent with $U/V$, and only the $B_{2g}$ and $A_{2u}$ states can be the ground states with their phase boundary always the same with that at $2U = V$. The phase boundary between the $B_{2g}$ and $A_{2u}$ states at $2U = V$ also applies to the condition with $U > 0$ and $V < 0$. It is worth pointing out that, all the states in the phase diagram are fully gapped. Especially, the $B_{2g}$ state is actually similar to the nodeless $d$-wave state in the iron-based superconductors\cite{PhysRevB.84.024529, Hirschfeld_2011, PhysRevLett.119.267001}.

A remarkable feature in the phase diagram in Fig.\ref{fig2} is that, the spin-triplet $A_{2u}$ state occupies a large area and it can be the ground state even in the weak SOC  limit. The phenomenon is closely related to the symmetry-enforced \textit{spin-sublattice-momentum lock} on the normal-state energy bands shown in Fig.\ref{fig1}. As analyzed, near the M point the sublattice-distinguished spin polarization lies in the $xy$ plane and the strength is proportional to $1/\sqrt{ 1 + t^{\prime 2} k^2 \sin^2 2\theta / 4 \lambda^2 }$ satisfying $\langle {\bf s}_B ({\bf k}) \rangle = \langle {\bf s}_A (-{\bf k}) \rangle$. In the small chemical potential condition, the spin on the Fermi surface is nearly fully polarized. When Cooper pair forms between two electrons with opposite momenta, the spin-sublattice-momentum lock in Fig.\ref{fig1} enforces the equal-spin pairing state with the spin polarized in the $xy$ plane in the intersublattice channel, which is exactly the $A_{2u}$ state in Fig.\ref{fig2}. Moreover, since the fully polarized spin near M is enforced by symmetries which is regardless of the strength of the SOC, the $A_{2u}$ state can appear as the ground state in the weak SOC condition as long as the chemical potential is small. In the large chemical potential condition, the average spin polarization on the large Fermi surface is weak. Correspondingly, the spin-singlet $A_{1g}$ and $B_{2g}$ states become more favorable. It is worth mentioning that in the limit $\mu \rightarrow 0$ and $\lambda \rightarrow 0$, the spin-singlet states compete with the spin triplet state, due to the concentric Fermi surface structure arising from the fourfold band degeneracy at M as indicated in Fig.\ref{fig1}.

\begin{figure}[!htbp]
	\centering
	\includegraphics[width=0.98\linewidth]{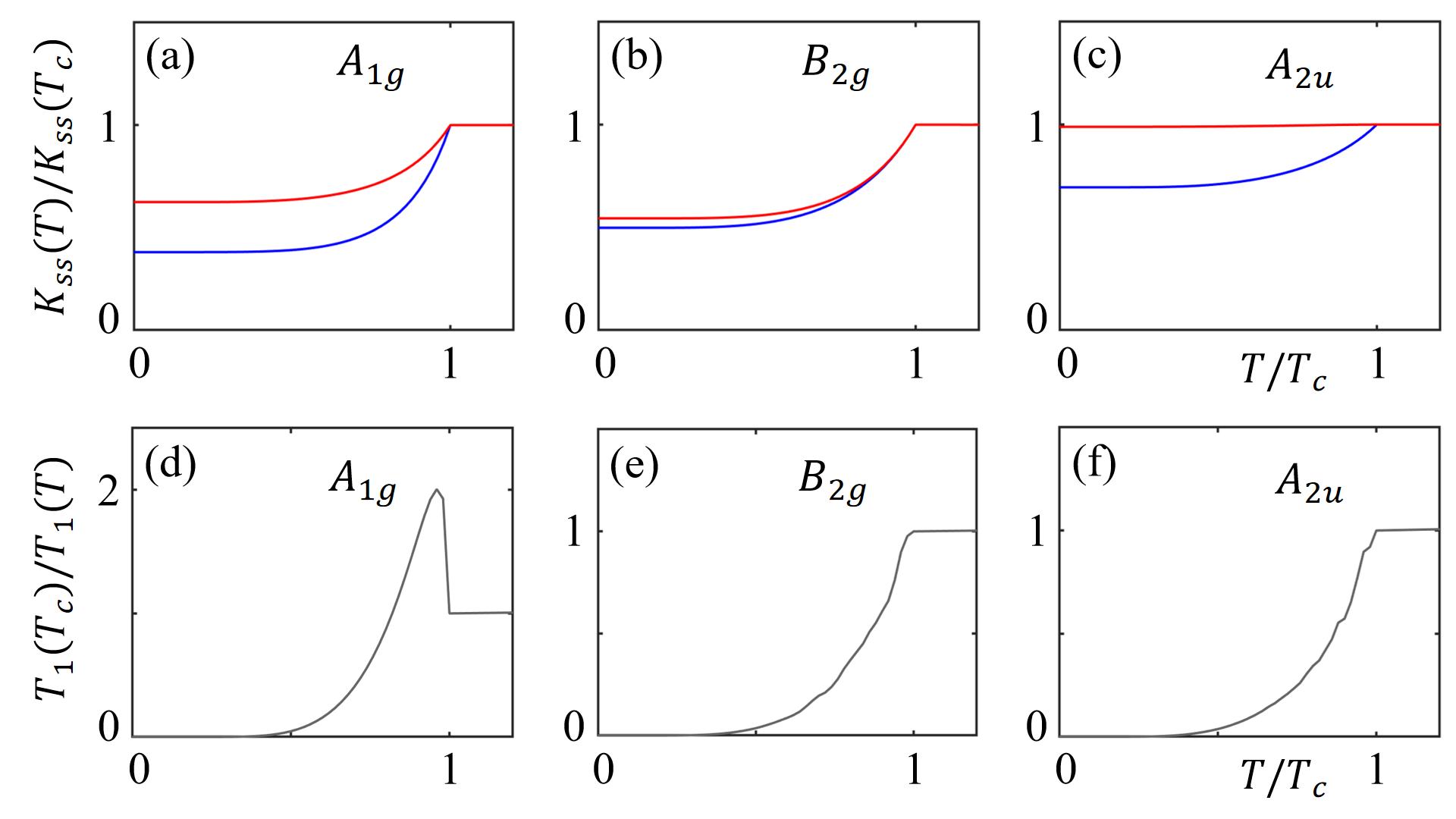}
	\caption{\label{fig3} (color online) (a)$\sim$(c) show the Knight shift and (d)$\sim$(f) show the spin relaxation rate versus the reduced temperature $T/T_c$, for the three possible superconducting ground states obtained in the phase diagram in Fig.\ref{fig2}. The red and blue lines in (a)$\sim$(c) correspond to the out-of-plane and in-plane Knight shift respectively. In the calculations, we set $\lambda = 0.2$ and the superconducting order $\Delta = 0.05$ for the $A_{1g}$ and $A_{2u}$ states, and $\{ \lambda, \Delta \} = \{ 0.1, 0.1 \}$ for the $B_{2g}$ state, in accordance with the phase diagram in Fig.\ref{fig2}. The other parameters are $\{ t, t^\prime, \mu \} = \{ 1.0, 0.8, 0.3 \}$. }
\end{figure}

\textit{Experimental signatures.}
The different states in the phase diagram in Fig.\ref{fig2} can be distinguished in experiments. In nuclear magnetic resonance measurements, the temperature dependence of the Knight shift $K_{ss}$ and the spin relaxation rate $1/T_1$   can provide essential information on the superconducting orders\cite{RevModPhys.75.657, RevModPhys.74.235, PhysRev.107.901, PhysRev.113.1504}. We calculate $K_{ss}$ and $1/T_1$ for the different superconducting ground states in Fig.\ref{fig2}. With a strong SOC, as shown in Fig.\ref{fig3}(a)$\sim$(c) the $A_{2u}$ state has a distinguishing feature in the Knight shift, i.e. the constant $K_{zz}$ corresponding to magnetic fields applied along the $z$ direction, and the $A_{1g}$ state is characterized by the Hebel Slichter coherence peak in the spin relaxation rate as presented in Fig.\ref{fig3}(d)$\sim$(f), as the temperature cools down below $T_c$. For the $B_{2g}$ state, the Knight shift is always suppressed and the Hebel Slichter coherence peak in the spin relaxation rate is absent. At  an ultra low temperature, all the three states show similar exponential scaling behavior in the spin relaxation rate as indicated in Fig.\ref{fig3}(d)$\sim$(f), due to their nodeless gap structures. We want to note that the Knight shift results can change if the strength of the SOC is comparable to the pairing order (more details in SM), while the features in the spin relaxation rate always hold for the different states.

Another characteristic feature for the $A_{2u}$ state is its in-plane upper critical field exceeding the Pauli limit, which is closely related to the following facts. (i) In the $A_{2u}$ state, the Cooper pair forms between an electron and its inversion partner, and the magnetic field preserves the inversion symmetry. (ii) The in-plane magnetic field only modifies the spin polarization in Fig.\ref{fig1} which is vital for the $A_{2u}$ state as analyzed in the above. For the $A_{1g}$ and $B_{2g}$ states, due to the spin-singlet nature, their in-plane upper critical fields obey the Pauli limit. In the SM, we roughly estimate the in-plane upper critical fields numerically. Notice that, here we omit the possible superconducting phase transitions, i.e. the phase transition from the even parity state to the odd parity state\cite{even_odd, PhysRevB.105.L020505} and the transition to the Fulde-Ferrell-Larkin-Ovchinnikov state\cite{PhysRev.135.A550, LO}, driven by the magnetic field; and we also ignore the symmetry breaking effect arising from the in-plane magnetic field.

In summary, we find that the nonsymmorphic lattice symmetries can greatly assist the spin-triplet superconductivity in the presence of SOC.  In a system respecting the space group $P4/nmm$, the nonsymmorphic symmetry makes the spin-triplet $A_{2u}$ state be the leading pairing instability because of the spin-sublattice-momentum lock on electronic bands. Topologically, the spin-triplet $A_{2u}$ state is trivial. The triviality can be easily understood from the concentric Fermi surface structure arising from the fourfold band degeneracy at M according to the parity criterion for centrosymmetric superconductors\cite{PhysRevLett.105.097001, PhysRevB.76.045302}. Our work unveils a new way in searching for the spin-triplet superconductors.


The authors are grateful to Xianxin Wu for fruitful discussions. This work is supported by the Ministry of Science and Technology of China 973 program (Grant No. 2017YFA0303100), National Science Foundation of China (Grant No. NSFC-12174428, NSFC-11888101 and NSFC-11920101005), and the Strategic Priority Research Program of Chinese Academy of Sciences (Grant No. XDB28000000  and No. XDB33000000).

\appendix
\onecolumngrid


\begin{widetext}

\renewcommand{\theequation}{A\arabic{equation}}
\renewcommand{\thefigure}{A\arabic{figure}}
\renewcommand{\thetable}{A\arabic{table}}
\setcounter{equation}{0}
\setcounter{figure}{0}
\setcounter{table}{0}

\section{Matrix form for the symmetry operators at G and M}
The nonsymmorphic symmetry must lead to multiple sublattices in the system, as indicated in the lattice structure respecting the space group $P4/nmm$ in the main text. Due to the sublattice degree of freedom, the matrix form for the symmetry operations is ${\bf k}$-dependent. In the following, we take the $\{ S_{4z} | {\bf 0} \}$ symmetry for instance and construct its matrix form at G and M. To do this, we need to figure out how the symmetry operations act on the basis $( | \phi_A ({\bf k}) \rangle, | \phi_B ({\bf k}) \rangle ) = ( c^\dagger_{{\bf k}}, d^\dagger_{{\bf k}} ) | 0 \rangle$, where $| 0 \rangle$ is the vacuum and the spin index has been omitted for convenience. The bases in the reciprocal space and in the real space are related by the Fourier transform
\begin{eqnarray}\label{SM_Fourier}
| \phi_A ({\bf k}) \rangle = \sum_j e^{i {\bf k} \cdot {\bf R}_A^j } | \phi_A ({\bf R}_A^j) \rangle, \quad | \phi_B ({\bf k}) \rangle = \sum_j e^{i {\bf k} \cdot {\bf R}_B^j } | \phi_B ({\bf R}_B^j) \rangle,
\end{eqnarray}
where ${\bf R}_{A}^j$ (${\bf R}_{B}^j$) labels the position of the A (B) site in the $jth$ unit cell and ${\bf R}_A^j - {\bf R}_B^j = {\bf \tau_0} = {\bf a}_1/2 + {\bf a}_2/2$ as shown in in the lattice in the main text. Under $\{ S_{4z} | {\bf 0} \}$, the G point is left unchanged, i.e. $g {\bf k} = {\bf k}$; however, the M point is transfromed as ${\bf k} \rightarrow {\bf k} - {\bf b}_1$. Accordingly, when $\{ S_{4z} | {\bf 0} \}$ acts on the basis function, we have
\begin{subequations}\label{SM_symmetry_transform}
\begin{align}
& {\text G}: \   \{ S_{4z} | {\bf 0} \} ( | \phi_A ({\bf k}) \rangle, | \phi_B ({\bf k}) \rangle )
= e^{i s_3 \pi/4} ( | \phi_A ({\bf k}) \rangle, | \phi_B ({\bf k}) \rangle )   \label{symmetry_transform1} \\
& {\text M}: \  \{ S_{4z} | {\bf 0} \} ( | \phi_A ({\bf k}) \rangle, | \phi_B ({\bf k}) \rangle )
= e^{i s_3 \pi/4} ( | \phi_A ({\bf k} - {\bf b}_1) \rangle, | \phi_B ({\bf k} - {\bf b}_1) \rangle ) \nonumber \\
& \quad = \sum_j e^{i s_3 \pi/4} ( e^{i ({\bf k} - {\bf b}_1) \cdot {\bf R}_A^j } | \phi_A ({\bf R}_A^j) \rangle,  e^{i ({\bf k} - {\bf b}_1) \cdot {\bf R}_A^j } e^{-i ({\bf k} - {\bf b}_1) \cdot {\bf \tau}_0 } | \phi_B ({\bf R}_B^j) \rangle ) \nonumber \\
& \quad = \sum_j e^{i s_3 \pi/4} ( e^{i {\bf k} \cdot {\bf R}_A^j } | \phi_A ({\bf R}_A^j) \rangle,  e^{i {\bf k} \cdot {\bf R}_A^j } e^{-i {\bf k} \cdot {\bf \tau}_0 } e^{i {\bf b}_1 \cdot {\bf \tau}_0 } | \phi_B ({\bf R}_B^j) \rangle ) \nonumber \\
& \quad = \sum_j e^{i s_3 \pi/4} ( e^{i {\bf k} \cdot {\bf R}_A^j } | \phi_A ({\bf R}_A^j) \rangle,  -e^{i {\bf k} \cdot {\bf R}_B^j } | \phi_B ({\bf R}_B^j) \rangle ) \nonumber \\
& \quad = e^{i s_3 \pi/4} ( | \phi_A ({\bf k}) \rangle, - | \phi_B ({\bf k}) \rangle ),
\label{symmetry_transform2}
\end{align}
\end{subequations}
where we have taken use of the fact that ${\bf R}_A^j \cdot {\bf b}_1$ $mod$ $2\pi$ equals 0 and ${\bf b}_1 \cdot {\bf \tau}_0 = \pi$. Therefore, in the normal state $\{ S_{4z} | {\bf 0} \}$ has the matrix form $e^{i s_3 \pi/4} \sigma_0$ at G and $e^{i s_3 \pi/4} \sigma_3$ at M. Similar analysis can be applied to other symmetry operations, and we get the results in Table.\ref{symmetry_matrix_SM}.

As pointed out in the main text, for the space group $P4/nmm$ in the spinful condition it merely has one 4D irreducible representation at M, which is contributed by states with angular momenta $J_z = \pm 1/2$ and $J_z = \pm 3/2$ defined according to $\{ S_{4z} | {\bf 0} \}$ (or the fourfold rotation $\{ C_{4z} | {\bf \tau}_0 \}$). However, in constructing the low-energy effective model in the main text we only consider the $s$ orbital. At first glance, the $s$ orbital can not contribute states beyond $J_z = \pm 1/2$. According to Eq.\eqref{symmetry_transform2}, the additional angular momentum origins from the plane wave part of the Bloch wave function.

\begin{table}[]
\caption{\label{symmetry_matrix_SM} { Matrix form for the symmetry operations at G and M. Here, $\mathcal{I}$, $\mathcal{M}_y$, $\mathcal{M}_{xy}$, $\mathcal{S}_{4z}$ and $\mathcal{T}$ standing for $\{ I | {\bf \tau_0} \}$, $\{ M_y | {\bf 0} \}$, $\{ M_{xy} | {\bf \tau_0} \}$, $\{ S_{4z} | {\bf 0} \}$ and the time reversal symmetry respectively. } }
\begin{tabular}{ p{1.0cm}<{\centering} p{1.8cm}<{\centering} p{1.8cm}<{\centering} p{3.6cm}<{\centering} p{2.2cm}<{\centering} p{2.0cm}<{\centering} }  \hline\hline
           &   $\mathcal{I}$   &  $\mathcal{M}_y$   &    $\mathcal{M}_{xy}$    &   $\mathcal{S}_{4z}$     &   $\mathcal{T}$      \\ \hline
G  &  $s_0 \sigma_1$  &  $i s_2 \sigma_0$  &    $i(s_1 + s_2) \sigma_1/\sqrt{2}$    &  $e^{ is_3\pi/4 } \sigma_0$   &  $i s_2 \sigma_0 K$    \\
M  &  $s_0 \sigma_1$  &  $i s_2 \sigma_3$  &    $i(s_1 + s_2) \sigma_1/\sqrt{2}$    &  $e^{ is_3\pi/4 } \sigma_3$   &  $i s_2 \sigma_0 K$     \\ \hline\hline
\end{tabular}
\end{table}

\renewcommand{\theequation}{B\arabic{equation}}
\renewcommand{\thefigure}{B\arabic{figure}}
\renewcommand{\thetable}{B\arabic{table}}
\setcounter{equation}{0}
\setcounter{figure}{0}
\setcounter{table}{0}

\section{Effective model at G}
In this section, we present the detailed construction of the low-energy effective model $\mathcal{H}_{\text eff, G}({\bf k})$ near the Brillouin zone center, i.e. the G point. The effective model near the M point shown in the main text can be constructed in a similar way.

We first consider the time reversal symmetry and the inversion symmetry, which constrain the system as
\begin{eqnarray}\label{SM_constrain_kp}
\mathcal{T} \mathcal{H}_{\text eff, G}({\bf k}) \mathcal{T}^{-1} = \mathcal{H}_{\text eff, G}(-{\bf k}), \quad
\mathcal{I} \mathcal{H}_{\text eff, G}({\bf k}) \mathcal{I}^{-1} = \mathcal{H}_{\text eff, G}(-{\bf k}).
\end{eqnarray}
The four-band model $\mathcal{H}_{\text eff, G}({\bf k})$ can be generally expressed in the form of the sixteen $\Gamma = s_i \sigma_j$ matrices. The constraints in Eq.\eqref{SM_constrain_kp} merely allow six $\Gamma$ matrices, i.e. $s_0 \sigma_0$, $s_0 \sigma_1$, $s_0 \sigma_2$, $s_1 \sigma_3$, $s_2 \sigma_3$ and $s_3 \sigma_3$, to appear in $\mathcal{H}_{\text eff, G}({\bf k})$. Then, we consider the constraints of the crystalline symmetries. Based on the matrix form of the symmetry operations in Table.\ref{symmetry_matrix_SM}, one can classify the above six matrices as shown in Table.\ref{SM_matrix_G}. Therefore, $\mathcal{H}_{\text eff, G}({\bf k})$ must take the following form
\begin{eqnarray}\label{SM_G_kp}
\mathcal{H}_{\text eff, G}({\bf k}) = m({\bf k}) s_0 \sigma_0 + t^\prime s_0 \sigma_1 + \lambda k_z s_2 \sigma_3 + \lambda k_y s_1 \sigma_3,
\end{eqnarray}
with $m({\bf k}) = t (k_x^2 + k_y^2)$ and $t, t^\prime, \lambda$ all the coefficients.

\begin{table}[]
\caption{\label{SM_matrix_G} Classification of the $s_i\sigma_j$ matrices which are allowed by the time reversal symmetry and the inversion symmetry, and the functions $f({\bf k})$ at G. The classification is according to the $D_{4h}$ point group. }
\begin{tabular}{ p{1.5cm}<{\centering} p{1.5cm}<{\centering} p{1.5cm}<{\centering} p{1.5cm}<{\centering} p{1.5cm}<{\centering} p{1.5cm}<{\centering} p{2.8cm}<{\centering} p{2.8cm}<{\centering} } \hline\hline
          &   $E$   &   $\mathcal{S}_{4z}$   &   $\mathcal{I}$   &   $\mathcal{M}_y$   &   $\mathcal{M}_{xy}$   &   space   &   $s_i \sigma_j$   \\ \hline
$A_{1g}$  &  1  &  1   &  1   &  1   &  1   &  $x^2 + y^2$  &  $s_0 \sigma_0$, $s_0 \sigma_1$  \\
$B_{1u}$  &  1  &  1   &  -1  &  -1  &  1   &               &  $s_3 \sigma_3$  \\
$B_{2u}$  &  1  &  1   &  -1  &  1   &  -1  &               &  $s_0 \sigma_2$  \\
$E_u$     &  2  &  0   &  -2  &  0   &  0   &  $(x, y)$     &  $( s_2 \sigma_3, s_1 \sigma_3)$  \\ \hline\hline
\end{tabular}
\end{table}

In fact, the effective model at G shown in Eq.\eqref{SM_G_kp} has similar form with the model Hamiltonian considered in Ref.\cite{PhysRevLett.108.147003}, but their physical meanings are different. However, if we consider the phenomenological density-density interactions similar to that in the main text, we can expect similar conclusions with that in Ref.\cite{PhysRevLett.108.147003} and the spin-triplet superconductivity can appear in the strong SOC  condition.

\renewcommand{\theequation}{C\arabic{equation}}
\renewcommand{\thefigure}{C\arabic{figure}}
\renewcommand{\thetable}{C\arabic{table}}
\setcounter{equation}{0}
\setcounter{figure}{0}
\setcounter{table}{0}

\section{Spin polarization on the energy bands}
In this part, we present the detailed calculations for the spin polarization on the energy bands based on the low-energy effective theory near the M point in the main text. The effective Hamiltonian near M can be solved as
\begin{eqnarray}\label{SM_solve_M}
| \varphi_1 ({\bf k}) \rangle &=& \frac{1}{ \sqrt{2} } \left(
	\begin{array}{c}
		-1    \\
		\cos\zeta   \\
        i\sin\zeta e^{-i\theta}   \\
        0
	\end{array}
	\right),   \qquad
| \varphi_2 ({\bf k}) \rangle = \frac{-1}{ \sqrt{2} } \left(
	\begin{array}{c}
		0    \\
        i\sin\zeta e^{i\theta}   \\
		\cos\zeta   \\
        -1
	\end{array}
	\right),       \nonumber \\
| \varphi_3 ({\bf k}) \rangle &=& \frac{1}{ \sqrt{2} } \left(
	\begin{array}{c}
		1    \\
		\cos\zeta   \\
        i\sin\zeta e^{-i\theta}   \\
        0
	\end{array}
	\right),    \qquad
| \varphi_4 ({\bf k}) \rangle = \frac{-1}{ \sqrt{2} } \left(
	\begin{array}{c}
		0    \\
        i\sin\zeta e^{i\theta}   \\
		\cos\zeta   \\
        1
	\end{array}
	\right),
\end{eqnarray}
where $\sin\zeta = \frac { \lambda k } { \sqrt{ \lambda^2 k^2 + t^{\prime 2} k^4 \sin^2 2\theta / 4 } }$ and $\cos\zeta = \frac { t^\prime k^2 \sin 2\theta } { 2 \sqrt{ \lambda^2 k^2 + t^{\prime 2} k^4 \sin^2 2\theta / 4 } }$ with ${\bf k}$ written in the polar coordinates $(k_x, k_y) = (k\sin\theta, k\cos\theta)$. In Eq.\eqref{SM_solve_M}, $| \varphi_1 ({\bf k}) \rangle$ and $| \varphi_2 ({\bf k}) \rangle$ are the two degenerate eigenstates corresponding to eigenvalue $E_- = tk^2 - \sqrt{ \lambda^2 k^2 + t^{\prime 2} k^4 \sin^2 2\theta / 4 }$, while $| \varphi_3 ({\bf k}) \rangle$ and $| \varphi_4 ({\bf k}) \rangle$ are the two degenerate eigenstates with eigenvalue $E_+ = tk^2 + \sqrt{ \lambda^2 k^2 + t^{\prime 2} k^4 \sin^2 2\theta / 4 }$.

The sublattice-distinguished spin operators are $s_{A/B, i} = s_i ( \sigma_0 \pm \sigma_3 ) / 2$. Straightforwardly, the spin polarization on the $E_-$ bands can be calculated as $\langle {\bf s}_{A/B} \rangle = \langle \varphi_1 | {\bf s}_{A/B} | \varphi_1 \rangle + \langle \varphi_2 | {\bf s}_{A/B} | \varphi_2 \rangle$, which turns out to be $\langle {\bf s}_B \rangle = -\langle {\bf s}_A \rangle = ( \sin\zeta\sin\theta, \sin\zeta\cos\theta, 0 )$. Obviously, $| \langle {\bf s}_{A/B} \rangle |$ at ${\bf k}$ is proportional to $\frac { 1 } { \sqrt{ 1 + t^{\prime 2} k^2 \sin^2 2\theta / 4 \lambda^2 } }$. The spin polarization on the $E_+$ bands can be calculated similarly.

For the Fermi surfaces near the G point, according to Eq.\eqref{SM_G_kp}, the spin polarization can be obtained as $| \langle {\bf s}_{A/B} \rangle | = \frac { 1 } { \sqrt{ 1 + t^{\prime 2} / 4 \lambda^2 k^2 } }$. Comparing the results near G and M, one immediately comes to the conclusion, the spin polarization is vanishing small for small Fermi surfaces near G, while it is nearly fully polarized for small Fermi surfaces near M.

\renewcommand{\theequation}{D\arabic{equation}}
\renewcommand{\thefigure}{D\arabic{figure}}
\renewcommand{\thetable}{D\arabic{table}}
\setcounter{equation}{0}
\setcounter{figure}{0}
\setcounter{table}{0}

\section{Derivation of the superconducting ground state}
In the main text, we get the superconducting ground states by solving the linearized gap equations. Here, we present the details on the derivation of the linearized gap equation, and present more analysis on the calculations of the superconducting susceptibility.

\subsection{Linearized gap equation}
In the superconducting state, the Green functions can be defined as
\begin{eqnarray}\label{Green_functions}
G_{ij}({\bf k}, \tau) &=& -\langle T_\tau c_i({\bf k}, \tau) c_j^\dagger({\bf k}, 0) \rangle,   \\
F_{ij}({\bf k}, \tau) &=& \langle T_\tau c_i({\bf k}, \tau) c_j(-{\bf k}, 0) \rangle,  \nonumber \\
F_{ij}^\dagger({\bf k}, \tau) &=& \langle T_\tau c_i^\dagger(-{\bf k}, \tau) c_j^\dagger({\bf k}, 0) \rangle.  \nonumber
\end{eqnarray}
In the mean-field level, the superconducting order can be calculated as
\begin{eqnarray}\label{order_Green}
{\bf \Delta({\bf k}) } = -\frac{1}{N} \sum_{{\bf k^\prime}} \mathcal{U}({\bf k, k^\prime}) F({\bf k^\prime}, \tau=0)
= \frac{1}{N\beta} \sum_{{\bf k^\prime}, n} \mathcal{U}({\bf k, k^\prime}) F({\bf k^\prime}, i\omega_n).
\end{eqnarray}
Notice that the Fourier transformation for Eq.\eqref{Green_functions} is as follows: $g(\tau) = \frac{1}{\beta} \sum_{n=-\infty}^{+\infty} e^{-i\omega_n\tau} g(i\omega_n)$ and $g(i\omega_n) = \int_0^\beta e^{i\omega_n\tau} g(\tau)$ with $\omega_n = \frac{2n\pi}{\beta}$ for boson and $\omega_n = \frac{(2n+1)\pi}{\beta}$ for fermion. According to Eq.\eqref{order_Green}, we must calculate $F({\bf k^\prime}, \tau=0)$ firstly. To do this, we consider the Gor'kov equations\cite{Gorkov_equation}, $i.e.$ the equation of motion in the superconducting state, which read as
\begin{eqnarray}\label{Gorkov_equation1}
&& \qquad G_0^{-1}({\bf k}, i\omega) G({\bf k}, i\omega) + {\bf \Delta}({\bf k}) F^\dagger({\bf k}, i\omega) = 1, \\
&& \quad G_0^{-1}({\bf k}, i\omega) F({\bf k}, i\omega) - {\bf \Delta}({\bf k}) G^T(-{\bf k}, -i\omega) = 0,  \nonumber \\
&&-(G_0^{-1})^T(-{\bf k}, -i\omega) F^\dagger({\bf k}, i\omega) + {\bf \Delta}^\dagger({\bf k}) G({\bf k}, i\omega) = 0.  \nonumber
\end{eqnarray}
According to the Gor'kov equations, we can derive the following equations
\begin{eqnarray}\label{Gorkov_equation2}
F^\dagger({\bf k}, i\omega) &=& G_0^T(-{\bf k}, -i\omega) {\bf \Delta}^\dagger({\bf k}) G({\bf k}, i\omega), \\
F({\bf k}, i\omega) &=& G_0({\bf k}, i\omega) {\bf \Delta}({\bf k}) G^T(-{\bf k}, -i\omega),  \nonumber \\
G^{-1}({\bf k}, i\omega) &=& G_0^{-1}({\bf k}, i\omega) + {\bf \Delta}({\bf k}) G_0^T(-{\bf k}, -i\omega) {\bf \Delta}^\dagger({\bf k}), \nonumber
\end{eqnarray}
where $G_0({\bf k}, i\omega)$ is the normal-state Green function. In the weak-coupling condition, ${\bf \Delta}$ is small and we have \begin{eqnarray}\label{Gorkov_equation3}
F({\bf k}, i\omega) & \simeq & G_0({\bf k}, i\omega) {\bf \Delta}({\bf k}) G_0^T(-{\bf k}, -i\omega)
= G_0({\bf k}, i\omega) {\bf \Delta}({\bf k}) G_0^\ast(-{\bf k}, i\omega),
\end{eqnarray}
where we have used the identity $G_0(-{\bf k}, -i\omega) = G_0^\dagger(-{\bf k}, i\omega)$. The normal-state Green function can be written in the band basis
\begin{eqnarray}\label{Green_normal_band}
G_0({\bf k}, i\omega) = \frac { 1 } { i\omega - h_0({\bf k}) }
= \sum_s \frac { | u_{s, {\bf k}} \rangle \langle u_{s, {\bf k}} | } { i\omega - \xi_{s, {\bf k}} },
\end{eqnarray}
where $s$ is the band index, and $| u_{s, {\bf k}} \rangle$ ($\xi_{s, {\bf k}}$)  is the eigenfunction (eigenenergy) for band $s$. Accordingly, the anomalous superconducting Green function is
\begin{eqnarray}\label{Green_SC_band1}
F({\bf k}, i\omega) & = & \sum_{ s,{s^\prime} } \frac { | u_{s, {\bf k}} \rangle \langle u_{s, {\bf k}} | {\bf \Delta}({\bf k}) | u^\ast_{{s^\prime}, {\bf -k}} \rangle \langle u^\ast_{{s^\prime}, {\bf -k}} | }  { ( i\omega - \xi_{s, {\bf k}} ) ( -i\omega - \xi_{{s^\prime}, {\bf -k}} ) },
\end{eqnarray}
where we use $| u^\ast_{{s^\prime}, {\bf -k}} \rangle$ to lable $( | u_{{s^\prime}, {\bf -k}} \rangle )^\ast$. In the weak-coupling condition, the superconductivity is mainly contributed by electrons on the Fermi surfaces. Moreover, since the system in our consideration possesses both the time reversal symmetry and inversion symmetry, the electron on the Fermi surfaces can always form Cooper pair with its time reversal or inversion partner, namely $s^\prime = \bar{s}$ or $s^\prime = s$, and
\begin{eqnarray}\label{Green_SC_band2}
F({\bf k}, i\omega) & = & \sum_{ s,{s^\prime} } \frac { | u_{s, {\bf k}} \rangle \langle u_{s, {\bf k}} | {\bf \Delta}({\bf k}) | u^\ast_{{s^\prime}, {\bf -k}} \rangle \langle u^\ast_{{s^\prime}, {\bf -k}} | }  { \omega^2 + \xi^2_{s, {\bf k}} }.
\end{eqnarray}
Correspondingly, the superconducting order parameter in Eq.\eqref{order_Green} is
\begin{eqnarray}\label{order_Green_cal}
{\bf \Delta({\bf k}) } &=& -\frac{1}{N\beta} \sum_{{\bf k^\prime}, n} \mathcal{U}({\bf k, k^\prime}) F({\bf k^\prime}, i\omega_n)
= -\frac{1}{N} \sum_{{\bf k^\prime}, s, s^\prime} \mathcal{U}({\bf k, k^\prime}) | u_{s, {\bf k^\prime}} \rangle \langle u_{s, {\bf k^\prime}} | {\bf \Delta}({\bf k^\prime}) | u^\ast_{{s^\prime}, {\bf -k^\prime}} \rangle \langle u^\ast_{{s^\prime}, {\bf -k^\prime}} |  \frac{1}{\beta} \sum_n \frac{1}{\omega_n^2 + \xi^2_{s, {\bf k^\prime}}}   \nonumber \\
&=& -\frac{1}{N} \sum_{{\bf k^\prime}, s} \frac{\mathcal{U}({\bf k, k^\prime})}{2\xi_{s, {\bf k^\prime} }} \tanh \frac{\beta\xi_{s, {\bf k^\prime} }}{2} \sum_{ s^\prime=s,\bar{s} } | u_{s, {\bf k^\prime}} \rangle \langle u_{s, {\bf k^\prime}} | {\bf \Delta}({\bf k^\prime}) | u^\ast_{{s^\prime}, {\bf -k^\prime}} \rangle \langle u^\ast_{{s^\prime}, {\bf -k^\prime}} |,
\end{eqnarray}
where only the electronic states on the Fermi surfaces are taken into account in the weak-coupling condition. In calculating the frequency summation in Eq.\eqref{order_Green_cal}, we have used the relation $\oint_{|z|\rightarrow\infty} \frac{dz}{2\pi i} \frac{1}{\xi^2-z^2} \frac{1}{e^{\beta z}+1} = 0$.

For each irreducible representation channel, the superconducting order parameter ${\bf \Delta(k)}$ can be expanded according to the corresponding bases. Considering the orthonormality of the basis functions, we have
\begin{eqnarray}\label{gap_equation}
\sum_{\bf k} {\text {tr}} [ {\bf \Delta}({\bf k}) \Delta^\dagger_{\nu^\prime}({\bf k}) ]
&=& \sum_{{\bf k}, \nu} \kappa_\nu {\text {tr}} [ \Delta_\nu({\bf k}) \Delta^\dagger_{\nu^\prime}({\bf k}) ] = \kappa_\nu   \\
&=& -\sum_{\bf k} {\text {tr}}  [  \frac{1}{N} \sum_{{\bf k^\prime}, s} \frac{\mathcal{U}({\bf k, k^\prime})}{2\xi_{s, {\bf k^\prime} }} \tanh \frac{\beta\xi_{s, {\bf k^\prime} }}{2} \sum_{ s^\prime=s,\bar{s} } | u_{s, {\bf k^\prime}} \rangle \langle u_{s, {\bf k^\prime}} | {\bf \Delta}({\bf k^\prime}) | u^\ast_{{s^\prime}, {\bf -k^\prime}} \rangle \langle u^\ast_{{s^\prime}, {\bf -k^\prime}} |  \Delta^\dagger_{\nu^\prime}({\bf k})  ]  \nonumber \\
&=& -\sum_\nu  \frac{1}{N} \sum_{{\bf k}, {\bf k^\prime}, s} \frac{\mathcal{U}({\bf k, k^\prime})}{2\xi_{s, {\bf k^\prime} }} \tanh \frac{\beta\xi_{s, {\bf k^\prime} }}{2} \sum_{ s^\prime=s,\bar{s} } \langle u_{s, {\bf k^\prime}} | \Delta_\nu({\bf k^\prime}) | u^\ast_{{s^\prime}, {\bf -k^\prime}} \rangle \langle u^\ast_{{s^\prime}, {\bf -k^\prime}} |  \Delta^\dagger_{\nu^\prime}({\bf k}) | u_{s, {\bf k^\prime}} \rangle  \kappa_\nu  \nonumber \\
&=& -\sum_\nu  \frac{1}{N} \sum_{{\bf k}, {\bf k^\prime}, s} \frac{\mathcal{U}({\bf k, k^\prime})}{2\xi_{s, {\bf k^\prime} }} \tanh \frac{\beta\xi_{s, {\bf k^\prime} }}{2} \sum_{ s^\prime=s,\bar{s} } \langle u_{s, {\bf k^\prime}} | \hat{\Delta}_\nu({\bf k^\prime}) | u_{{s^\prime}, {\bf k^\prime}} \rangle \langle u_{{s^\prime}, {\bf k^\prime}} |  \hat{\Delta}^\dagger_{\nu^\prime}({\bf k}) | u_{s, {\bf k^\prime}} \rangle  \kappa_\nu,  \nonumber
\end{eqnarray}
where $\Delta_\nu({\bf k}) = \hat{\Delta}_\nu({\bf k}) \cdot is_2$. Eq.\eqref{gap_equation} can be intuitively expressed in the form ${\bf \kappa} \cdot X = {\bf \kappa}$ with
\begin{eqnarray}\label{gap_equation2}
X_{\nu, \nu^\prime}(\beta) = -\frac{1}{N} \sum_{{\bf k}, {\bf k^\prime}, s} \frac{\mathcal{U}({\bf k, k^\prime})}{2\xi_{s, {\bf k^\prime} }} \tanh \frac{\beta\xi_{s, {\bf k^\prime} }}{2} \sum_{ s^\prime=s,\bar{s} }  \langle u_{s, {\bf k^\prime}} | \hat{\Delta}_\nu({\bf k^\prime}) | u_{{s^\prime}, {\bf k^\prime}} \rangle \langle u_{{s^\prime}, {\bf k^\prime}} |  \hat{\Delta}^\dagger_{\nu^\prime}({\bf k}) | u_{s, {\bf k^\prime}} \rangle.
\end{eqnarray}
In our consideration the interaction is ${\bf k}$-independent, and in the continuum condition we have
\begin{eqnarray}\label{gap_equation2_kp}
X_{\nu, \nu^\prime}(\beta) &=& -\frac{1}{N} \sum_s \int d{\bf k^\prime} \frac{\mathcal{U}}{2\xi_{s, {\bf k^\prime} }} \tanh \frac{\beta\xi_{s, {\bf k^\prime} }}{2} \sum_{ s^\prime=s,\bar{s} }  \langle u_{s, {\bf k^\prime}} | \hat{\Delta}_\nu | u_{{s^\prime}, {\bf k^\prime}} \rangle \langle u_{{s^\prime}, {\bf k^\prime}} |  \hat{\Delta}^\dagger_{\nu^\prime} | u_{s, {\bf k^\prime}} \rangle, \\
&=& -\frac{1}{N} \sum_s \int \frac{dk^\prime}{d\xi_{s, {\bf k^\prime} }} d\xi_{s, {\bf k^\prime} } d\theta \frac{\mathcal{U}}{2\xi_{s, {\bf k^\prime} }} \tanh \frac{\beta\xi_{s, {\bf k^\prime} }}{2} \sum_{ s^\prime=s,\bar{s} }  \langle u_{s, {\bf k^\prime}} | \hat{\Delta}_\nu | u_{{s^\prime}, {\bf k^\prime}} \rangle \langle u_{{s^\prime}, {\bf k^\prime}} |  \hat{\Delta}^\dagger_{\nu^\prime} | u_{s, {\bf k^\prime}} \rangle,  \nonumber \\
&=& -\frac{\mathcal{U}}{N} \sum_s \int^{\omega_0}_{-\omega_0} d\xi_{s, {\bf k^\prime} } \frac{1}{2\xi_{s, {\bf k^\prime} }} \tanh \frac{\beta\xi_{s, {\bf k^\prime} }}{2}  \int d\theta D(\theta) \sum_{ s^\prime=s,\bar{s} }  \langle u_{s, {\bf k^\prime}} | \hat{\Delta}_\nu | u_{{s^\prime}, {\bf k^\prime}} \rangle \langle u_{{s^\prime}, {\bf k^\prime}} |  \hat{\Delta}^\dagger_{\nu^\prime} | u_{s, {\bf k^\prime}} \rangle,  \nonumber \\
&=& -\mathcal{U} \mathcal{F}(\beta) \sum_s \int d\theta D(\theta) \sum_{ s^\prime=s,\bar{s} }  \langle u_{s, {\bf k^\prime}} | \hat{\Delta}_\nu | u_{{s^\prime}, {\bf k^\prime}} \rangle \langle u_{{s^\prime}, {\bf k^\prime}} |  \hat{\Delta}^\dagger_{\nu^\prime} | u_{s, {\bf k^\prime}} \rangle
= -\mathcal{U} \chi_{\nu, \nu^\prime} (T_c).  \nonumber
\end{eqnarray}
In Eq.\eqref{gap_equation2_kp}, $\mathcal{F}(\beta) = \frac{1}{2N} \int^{\omega_0}_{-\omega_0}  \frac{1}{2\xi_{s, {\bf k^\prime} }} \tanh \frac{\beta\xi_{s, {\bf k^\prime} }}{2}  d\xi_{s, {\bf k^\prime} }$ is a temperature-dependent constant with $\omega_0$ the energy cutoff near the Fermi energy, and $D(\theta) = 2dk^\prime / d\xi_{s, {\bf k^\prime} }$ is the density of states on the Fermi surface. By solving the characteristic equation $-\mathcal{U} \chi (T_c) = I$, we can get the superconducting ground state.

\subsection{Superconductivity from density-density interactions}

In the main text, we consider superconductivity induced from the phenomenological density-density interactions
\begin{eqnarray}\label{SM_interaction_kp}
\mathcal{H}_{int} &=& \int d{\bf q} [ U \sum_{i=1}^2 n_i({\bf q}) n_i(-{\bf q}) + 2V n_1({\bf q}) n_2(-{\bf q}) ], \\
&=& \frac{1}{N} \int d{\bf k} d{\bf k^\prime} d{\bf q} ( U c^\dagger_{{\bf k^\prime+q},\sigma} c_{{\bf k^\prime},\sigma} c^\dagger_{{\bf k-q},{\bar{\sigma}}} c_{{\bf k},{\bar{\sigma}}} + U d^\dagger_{{\bf k^\prime+q},\sigma} d_{{\bf k^\prime},\sigma} d^\dagger_{{\bf k-q},{\bar{\sigma}}} d_{{\bf k},{\bar{\sigma}}}
+ 2V c^\dagger_{{\bf k^\prime+q},\sigma} c_{{\bf k^\prime},\sigma} d^\dagger_{{\bf k-q},{\sigma^\prime}} d_{{\bf k},{\sigma^\prime}} ).  \nonumber
\end{eqnarray}
In the superconducting channel, i.e. ${\bf k} = -{\bf k^\prime}$, we have
\begin{eqnarray}\label{SM_interaction_sc_kp}
\mathcal{H}_{int} &=& \frac{1}{N} \int d{\bf k^\prime} d{\bf q} ( U c^\dagger_{{\bf k^\prime+q},\sigma} c_{{\bf k^\prime},\sigma} c^\dagger_{{\bf -k^\prime-q},{\bar{\sigma}}} c_{{\bf -k^\prime},{\bar{\sigma}}} + U d^\dagger_{{\bf k^\prime+q},\sigma} d_{{\bf k^\prime},\sigma} d^\dagger_{{\bf -k^\prime-q},{\bar{\sigma}}} d_{{\bf -k^\prime},{\bar{\sigma}}}
+ 2V c^\dagger_{{\bf k^\prime+q},\sigma} c_{{\bf k^\prime},\sigma} d^\dagger_{{\bf -k^\prime-q},{\sigma^\prime}} d_{{\bf -k^\prime},{\sigma^\prime}} )  \nonumber \\
&=& \frac{1}{N} \int d{\bf k} d{\bf k^\prime} ( U c^\dagger_{{\bf k},\sigma} c^\dagger_{{\bf -k},{\bar{\sigma}}} c_{{\bf -k^\prime},{\bar{\sigma}}} c_{{\bf k^\prime},\sigma} + U d^\dagger_{{\bf k},\sigma} d^\dagger_{{\bf -k},{\bar{\sigma}}} d_{{\bf -k^\prime},{\bar{\sigma}}} d_{{\bf k^\prime},\sigma}
+ 2V c^\dagger_{{\bf k},\sigma} d^\dagger_{{\bf -k},{\sigma^\prime}} d_{{\bf -k^\prime},{\sigma^\prime}} c_{{\bf k^\prime},\sigma} ).
\end{eqnarray}
The interactions in Eq.\eqref{SM_interaction_sc_kp} can be expanded according to the superconducting orders. When we consider Fermi surfaces near the M point, we need to expand the interactions according to the pairing orders classified at M which is shown in the main text
\begin{eqnarray}\label{SM_interaction_sc_decouple_M}
\mathcal{H}_{int, M} &=& \frac{1}{N} \int d{\bf k} d{\bf k^\prime} [ U ( c^\dagger_{{\bf k},\sigma} c^\dagger_{{\bf -k},{\bar{\sigma}}} c_{{\bf -k^\prime},{\bar{\sigma}}} c_{{\bf k^\prime},\sigma} + d^\dagger_{{\bf k},\sigma} d^\dagger_{{\bf -k},{\bar{\sigma}}} d_{{\bf -k^\prime},{\bar{\sigma}}} d_{{\bf k^\prime},\sigma} )
+ 2V c^\dagger_{{\bf k},\sigma} d^\dagger_{{\bf -k},{\sigma^\prime}} d_{{\bf -k^\prime},{\sigma^\prime}} c_{{\bf k^\prime},\sigma} ]  \\
&=& \frac{1}{4N} \int d{\bf k} d{\bf k^\prime} [ U ( \hat{\Delta}_{ A_{1g} } \hat{\Delta}_{ A_{1g} }^\dagger + \hat{\Delta}_{ B_{2u} } \hat{\Delta}_{ B_{2u} }^\dagger ) + V ( \hat{\Delta}_{ A_{2u} } \hat{\Delta}_{ A_{2u} }^\dagger + \hat{\Delta}_{ B_{2g} } \hat{\Delta}_{ B_{2g} }^\dagger + \hat{\Delta}_{ E_u }^{(1)} \hat{\Delta}_{ E_u }^{(1) \dagger} + \hat{\Delta}_{ E_u }^{(2)} \hat{\Delta}_{ E_u }^{(2) \dagger} ) ].  \nonumber
\end{eqnarray}
Based on Eqs.\eqref{gap_equation2_kp}\eqref{SM_interaction_sc_decouple_M}, we can ge the linearized gap equations for each irreducible representation channel shown in the main text.

\subsection{Calculations of the superconducting susceptibility}
Based on the wave functions in Eq.\eqref{SM_solve_M}, we can calculate the superconducting susceptibility for each irreducible representation channel at M shown in the main text straightforwardly
\begin{eqnarray}\label{SM_superconductor_susceptibility}
\chi_{A_{1g}} &=& \mathcal{F}(\beta) \sum_s \sum_{ s^\prime=s,\bar{s} } \int d\theta [ D_-(\theta)  | \langle u_{s, {\bf k^\prime}}^- | s_0\sigma_0 | u_{{s^\prime}, {\bf k^\prime}}^- \rangle |^2 + D_+(\theta)  | \langle u_{s, {\bf k^\prime}}^+ | s_0\sigma_0 | u_{{s^\prime}, {\bf k^\prime}}^+ \rangle |^2 ]
= \mathcal{F}(\beta) \int d\theta [ D_-(\theta) + D_+(\theta) ]2,   \nonumber \\
\chi_{B_{2g}} &=& \mathcal{F}(\beta) \int d\theta [ D_-(\theta) + D_+(\theta) ] 2\cos^2\zeta,   \qquad \qquad
\chi_{A_{2u}} = \mathcal{F}(\beta) \int d\theta [ D_-(\theta) + D_+(\theta) ] 2\sin^2\zeta,   \nonumber \\
\chi_{B_{2u}} &=& \mathcal{F}(\beta) \int d\theta [ D_-(\theta) + D_+(\theta) ] 2\sin^2\zeta,   \qquad \qquad
\chi_{E_{u}} = \mathcal{F}(\beta) \int d\theta [ D_-(\theta) + D_+(\theta) ] 2\sin^2\zeta \sin^2\theta.
\end{eqnarray}
In the above equations, $D_\pm$ and $| u_{{s^\prime}}^\pm \rangle$ are the density of states and the eigenstates on the Fermi surface contributed by the energy band $E_\pm$ respectively. According to the results in Eq.\eqref{SM_superconductor_susceptibility} it is obvious to notice that, in the intrasublattice pairing channels $\chi_{A_{1g}} > \chi_{B_{2u}}$ and in the intersublattice channels $\chi_{A_{2u}} > \chi_{E_{u}}$. Moreover, considering that $\chi_{A_{2u}} + \chi_{B_{2g}} = \chi_{A_{1g}}$, if we compare the $A_{1g}$ state with a special case in the intersublattice pairing channels where $\chi_{A_{2u}} = \chi_{B_{2g}}$, i.e. the phase boundary between the $A_{2u}$ and $B_{2g}$ state, one can find that the $A_{1g}$ state can never be the ground state for $2|U| < |V|$ assuming $V$ attractive, which is consistent with the phase diagram in the main text.

\renewcommand{\theequation}{E\arabic{equation}}
\renewcommand{\thefigure}{E\arabic{figure}}
\renewcommand{\thetable}{E\arabic{table}}
\setcounter{equation}{0}
\setcounter{figure}{0}
\setcounter{table}{0}

\section{Lattice Model}
In the main text, we analyze the ${\bf k}$-independent superconducting ground states based on a low-energy effective model. Here, we show that these ${\bf k}$-independent states are the leading order approximation, i.e. the superconducting orders preserved to the $0th$ order of ${\bf k}$, in the minimal lattice model.

Taking the lattice structure in the main text into consideration and substituting $k_{x/y}$ with appropriate trigonometric functions, we can get the corresponding lattice model
\begin{eqnarray}\label{SM_lattice_Hamiltonian}
\mathcal{H} ({\bf k}) &=& 2t (\cos k_x + \cos k_y + 2) s_0\sigma_0 - \lambda \sin k_x s_2\sigma_3 - \lambda \sin k_y s_1\sigma_3 + 4t^\prime \cos\frac{k_x}{2} \cos\frac{k_y}{2} s_0\sigma_1.
\end{eqnarray}
The above lattice model respects the space group $P4/nmm$ with the lattice sites located at the $D_{2d}$ invariant points, and at each lattice site an $s$ orbital is considered as pointed out in the main text. In Eq.\eqref{SM_lattice_Hamiltonian}, $t$ is the intrasublattice nearest-neighbour hopping, $t^\prime$ is the intersublattice nearest-neighbour hopping, and $\lambda/2$ is the inversion-symmetric Rashba SOC. Based on the lattice model, we calculate the sublattice-distinguished spin polarization on the energy bands and plot the results in Fig.\ref{SM_spin}. As shown, the lattice model captures the essential features analyzed in the main text and in the above sections: for a system respecting the space group $P4/nmm$, it has fully polarized sublattice-distinguished spin polarization on the energy bands near the Brillouin zone boundary, i.e. $k_{x/y} = \pi$; while the spin polarization is vanishing small near the Brillouin zone center.

\begin{figure}[!htbp]
	\centering
	\includegraphics[width=0.5\linewidth]{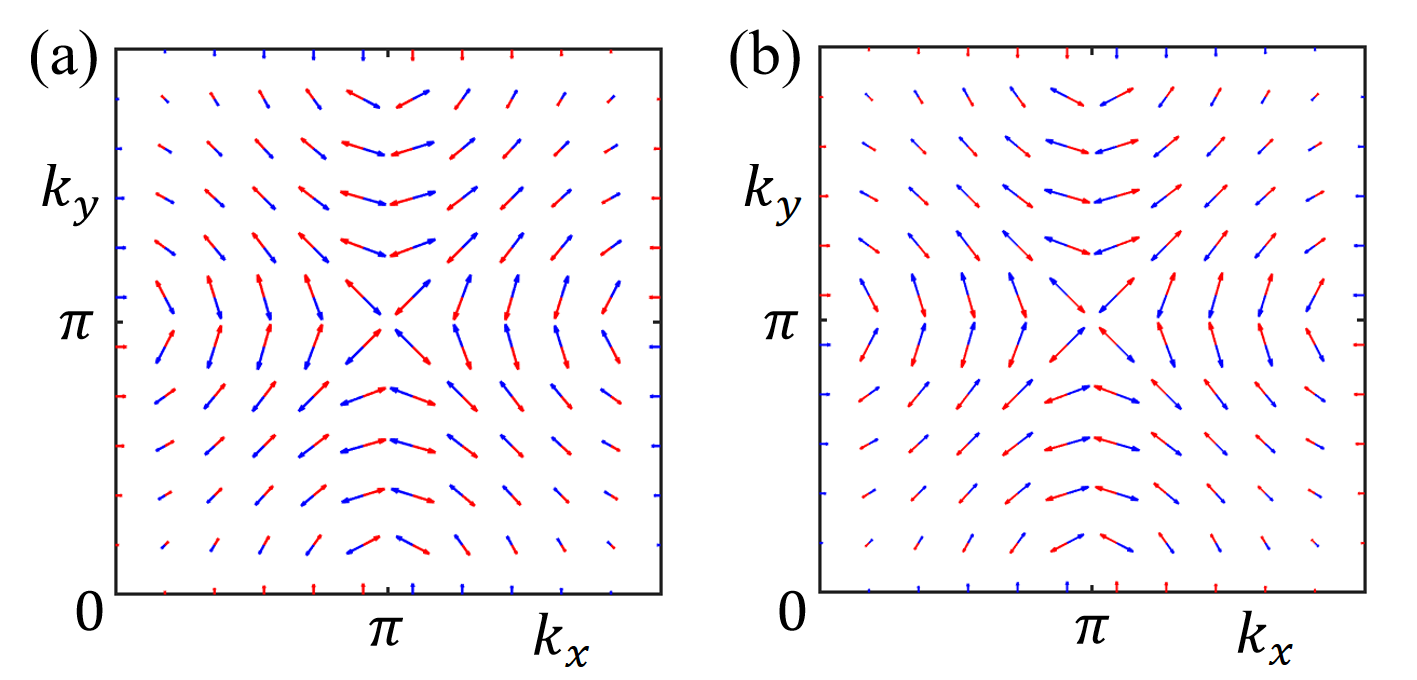}
	\caption{\label{SM_spin} (color online) Sketch for the sublattice-distinguished spin polarizations on the lower energy bands (a) and upper energy bands (b), plotted from the lattice model in Eq.\eqref{SM_lattice_Hamiltonian}. The parameters are set to be $\{ t, t^\prime, \lambda \} = \{ -1.0, 0.8, 0.8 \}$. The symbols in the figures are the same with that in the main text. }
\end{figure}

For the interacting part, we consider the following electron density-density interaction
\begin{eqnarray}\label{SM_interaction_Hamiltonian_real}
\mathcal{H}_{int}({\bf k}) &=& U \sum_{i,\sigma} n_{i\sigma} n_{i\bar{\sigma}} + V \sum_{\langle ij \rangle, \sigma, \sigma^\prime} n_{i\sigma} n_{j\sigma^\prime},
\end{eqnarray}
where $U$ is the onsite interaction and $V$ is the intersublattice nearest-neighbour interaction. By doing a Fourier transformation, $c_i = \frac{1}{\sqrt{N}} \sum\limits_{{\bf k}} c_{{\bf k}} e^{i {\bf k} \cdot {\bf r}_i}$ and $\frac{1}{N} \sum\limits_i e^{i ({\bf k - k^\prime}) \cdot {\bf r}_i} = \delta_{{\bf k, k^\prime}}$, we obtain
\begin{eqnarray}\label{SM_interaction_Hamiltonian_k}
\mathcal{H}_{int}({\bf k}) &=& \frac{U}{N} \sum_{{\bf k k^\prime q}} ( c^\dagger_{{\bf k^\prime+q},\sigma} c_{{\bf k^\prime},\sigma} c^\dagger_{{\bf k-q},\bar{\sigma}} c_{{\bf k},\bar{\sigma}} + d^\dagger_{{\bf k^\prime+q},\sigma} d_{{\bf k^\prime},\sigma} d^\dagger_{{\bf k-q},\bar{\sigma}} d_{{\bf k},\bar{\sigma}} )
+ \frac{V}{N} \sum_{{\bf k k^\prime q}} c^\dagger_{{\bf k^\prime+q},\sigma} c_{{\bf k^\prime},\sigma} d^\dagger_{{\bf k-q},\sigma^\prime} d_{{\bf k},\sigma^\prime} \sum_{{\bf \delta}=\langle ij \rangle} e^{i {{\bf q \cdot ({\bf r}_i - {\bf r}_j) }} } \nonumber \\
&=& \frac{U}{N} \sum_{{\bf k k^\prime q}} ( c^\dagger_{{\bf k^\prime+q},\sigma} c_{{\bf k^\prime},\sigma} c^\dagger_{{\bf k-q},\bar{\sigma}} c_{{\bf k},\bar{\sigma}} + d^\dagger_{{\bf k^\prime+q},\sigma} d_{{\bf k^\prime},\sigma} d^\dagger_{{\bf k-q},\bar{\sigma}} d_{{\bf k},\bar{\sigma}} )
+ \frac{V}{N} \sum_{{\bf k k^\prime q}} c^\dagger_{{\bf k^\prime+q},\sigma} c_{{\bf k^\prime},\sigma} d^\dagger_{{\bf k-q},\sigma^\prime} d_{{\bf k},\sigma^\prime} 4\cos\frac{q_x}{2} \cos\frac{q_y}{2}. \nonumber \\
\end{eqnarray}
In the superconducting channel, we set ${\bf k} = -{\bf k^\prime}$ and get
\begin{eqnarray}\label{SM_interaction_Hamiltonian_sc}
\mathcal{H}_{int} ({\bf k}) &=& \frac{U}{N} \sum_{{\bf k^\prime q}} ( c^\dagger_{{\bf k^\prime+q},\sigma} c_{{\bf k^\prime},\sigma} c^\dagger_{{\bf -k^\prime-q},\bar{\sigma}} c_{{\bf -k^\prime},\bar{\sigma}} + d^\dagger_{{\bf k^\prime+q},\sigma} d_{{\bf k^\prime},\sigma} d^\dagger_{{\bf -k^\prime-q},\bar{\sigma}} d_{{\bf -k^\prime},\bar{\sigma}} )  \nonumber \\
&& + \frac{V}{N} \sum_{{\bf k^\prime q}} c^\dagger_{{\bf k^\prime+q},\sigma} c_{{\bf k^\prime},\sigma} d^\dagger_{{\bf -k^\prime-q},\sigma^\prime} d_{{\bf -k^\prime},\sigma^\prime} 4\cos\frac{q_x}{2} \cos\frac{q_y}{2}  \nonumber \\
&=& \frac{U}{N} \sum_{{\bf k k^\prime}} ( c^\dagger_{{\bf k},\sigma} c^\dagger_{{\bf -k},\bar{\sigma}} c_{{\bf -k^\prime},\bar{\sigma}} c_{{\bf k^\prime},\sigma} + d^\dagger_{{\bf k},\sigma} d^\dagger_{{\bf -k},\bar{\sigma}} d_{{\bf -k^\prime},\bar{\sigma}} d_{{\bf k^\prime},\sigma})  \nonumber \\
&& + \frac{V}{N} \sum_{{\bf k k^\prime}} c^\dagger_{{\bf k},\sigma} d^\dagger_{{\bf -k},\sigma^\prime} d_{{\bf -k^\prime},\sigma^\prime} c_{{\bf k^\prime},\sigma} 4\cos\frac{k_x - k_x^\prime}{2} \cos\frac{k_y - k_y^\prime}{2}.
\end{eqnarray}
Apparently, the onsite interaction can only contribute the constant intrasublattice pairing order. For the intersublattice part, we have
\begin{eqnarray}\label{SM_interaction_Hamiltonian_NN}
\mathcal{H}_{int, NN}({\bf k}) &=& \frac{V}{N} \sum_{{\bf k k^\prime}} c^\dagger_{{\bf k},\sigma} d^\dagger_{{\bf -k},\sigma^\prime} d_{{\bf -k^\prime},\sigma^\prime} c_{{\bf k^\prime},\sigma} ( \cos\frac{k_x}{2} \cos\frac{k_y}{2} \cos\frac{k_x^\prime}{2} \cos\frac{k_y^\prime}{2} + \sin\frac{k_x}{2} \sin\frac{k_y}{2} \sin\frac{k_x^\prime}{2} \sin\frac{k_y^\prime}{2}  \nonumber \\
&& + \cos\frac{k_x}{2} \sin\frac{k_y}{2} \cos\frac{k_x^\prime}{2} \sin\frac{k_y^\prime}{2} + \sin\frac{k_x}{2} \cos\frac{k_y}{2} \sin\frac{k_x^\prime}{2} \cos\frac{k_y^\prime}{2} ).
\end{eqnarray}
According to the equation, it can be noticed that the term with form factor $\sin\frac{k_x}{2} \sin\frac{k_y}{2} \sin\frac{k_x^\prime}{2} \sin\frac{k_y^\prime}{2}$ dominates the other terms for small Fermi surfaces near the M point ($k_{x/y} \sim \pi$), since $\sin\frac{k_{x/y}}{2} = 1 - \frac{k_{M,x/y}^2}{2}$ and $\cos\frac{k_{x/y}}{2} = - \frac{k_{M,x/y}}{2}$ with ${\bf k}_M = {\bf k} - {\bf K}_M$ and ${\bf K}_M$ being the M point (${\bf k}_M$ is samll). Therefore, for small Fermi surfaces near M, it is reasonable we only consider the constant pairing orders between different sublattices, i.e. the $0th$ order of ${\bf k}$-dependent pairing orders contributed by the $\sin\frac{k_x}{2} \sin\frac{k_y}{2} \sin\frac{k_x^\prime}{2} \sin\frac{k_y^\prime}{2}$ term in Eq.\eqref{SM_interaction_Hamiltonian_NN}, which is exactly the consideration in the main text.

\renewcommand{\theequation}{F\arabic{equation}}
\renewcommand{\thefigure}{F\arabic{figure}}
\renewcommand{\thetable}{F\arabic{table}}
\setcounter{equation}{0}
\setcounter{figure}{0}
\setcounter{table}{0}

\section{Magnetic response}
In this part, we provide more details on the numerical simulation for the Knight shift and the spin relaxation rate, and present a rough numerical estimation for the in-plane upper critical field, for the superconducting ground states in the phase diagram in the main text.

\subsection{Knight shift and spin relaxation rate}
In the nuclear magnetic resonance, the Knight shift $K_{ss}$ and the spin relaxation rate $1/T_1$ are measured through the static spin susceptibility. In the general condition, the spin susceptibility is defined as
\begin{eqnarray}\label{SM_spin_susceptibility}
\chi_{st}( {\bf q}, i\omega) = \int_0^\beta d\tau \chi_{st}( {\bf q}, \tau)
= \int_0^\beta d\tau \langle T_\tau S_s({\bf q}) S_t(-{\bf q}) \rangle e^{i\omega\tau}.
\end{eqnarray}
In our consideration, the Knight shift in the nuclear magnetic resonance reads
\begin{eqnarray}\label{SM_Knight}
K_{ss}(T) \propto \sum_{\alpha} \chi_{ss}^{\alpha\alpha}(0,0)
\propto -\sum_{ {\bf k}, m, n, \alpha }  | \langle \phi_m ({\bf k}) | S_s^\alpha | \phi_n ({\bf k}) \rangle |^2 \frac{ n(E_{{\bf k} m}) - n(E_{{\bf k} n}) } { E_{{\bf k} m} - E_{{\bf k} n} }.
\end{eqnarray}
The spin relaxation rate is
\begin{eqnarray}\label{SM_spin_relax}
\frac {1} {T_1(T)} & \propto & \lim_{\omega\rightarrow 0} \sum_{{\bf q}, \alpha, s}  | A({\bf q}) |^2 \frac{ Im \chi_{ss}^{\alpha\alpha}({\bf q}, \omega+i0^+) } { \omega }    \\
& \propto & -\sum_{ {\bf k}, {\bf k}^\prime, m, n, s, \alpha }  | A({\bf k - k^\prime}) |^2  | \langle \phi_m ({\bf k}) | S_s^\alpha | \phi_n ({\bf k}^\prime) \rangle |^2 \frac{ \partial n(E) } { \partial E } \bigg|_{E = E_{{\bf k} m}} \delta( E_{{\bf k} m} - E_{{\bf k}^\prime n} ).  \nonumber
\end{eqnarray}
In the above equations, $S_s^\alpha$ ($s = 1, 2, 3$ and $\alpha = A, B$) is the spin operator for sublattice $\alpha$ in the Nambu space $(\psi^\dag({\bf k}), is_2 \psi(-{\bf k}))$, with $\psi^\dag({\bf k})$ being the basis for the normal-state Hamiltonian shown in the main text. Specifically, $S_s^A = s_s \otimes \frac{\sigma_0 + \sigma_3}{2} \otimes \tau_0$ and $S_s^B = s_s \otimes \frac{\sigma_0 - \sigma_3}{2} \otimes \tau_0$, with $\tau_0 = I_{2\times 2}$ in the Nambu space. $E_{{\bf k} m}$ and $\phi_m ({\bf k})$ are the energy and wavefunction for the $mth$ eigenstate for the superconducting Hamiltonian respectively, and $A({\bf q})$ in Eq.\eqref{SM_spin_relax} is the structure factor which is set to be $1$. In the numerical calculations, we set the superconducting transition temperature $k_bT_c = \Delta_0/3.53$ with $\Delta_0$ the zero temperature pairing order, and consider the $T$-dependent superconducting order $\Delta(T) = \Delta_0 f(T/T_c)$ with $f(T/T_c)$ being the BCS-type normalized gap presented in Ref.\cite{BCS_gap}.

\begin{figure}[!htbp]
	\centering
	\includegraphics[width=0.9\linewidth]{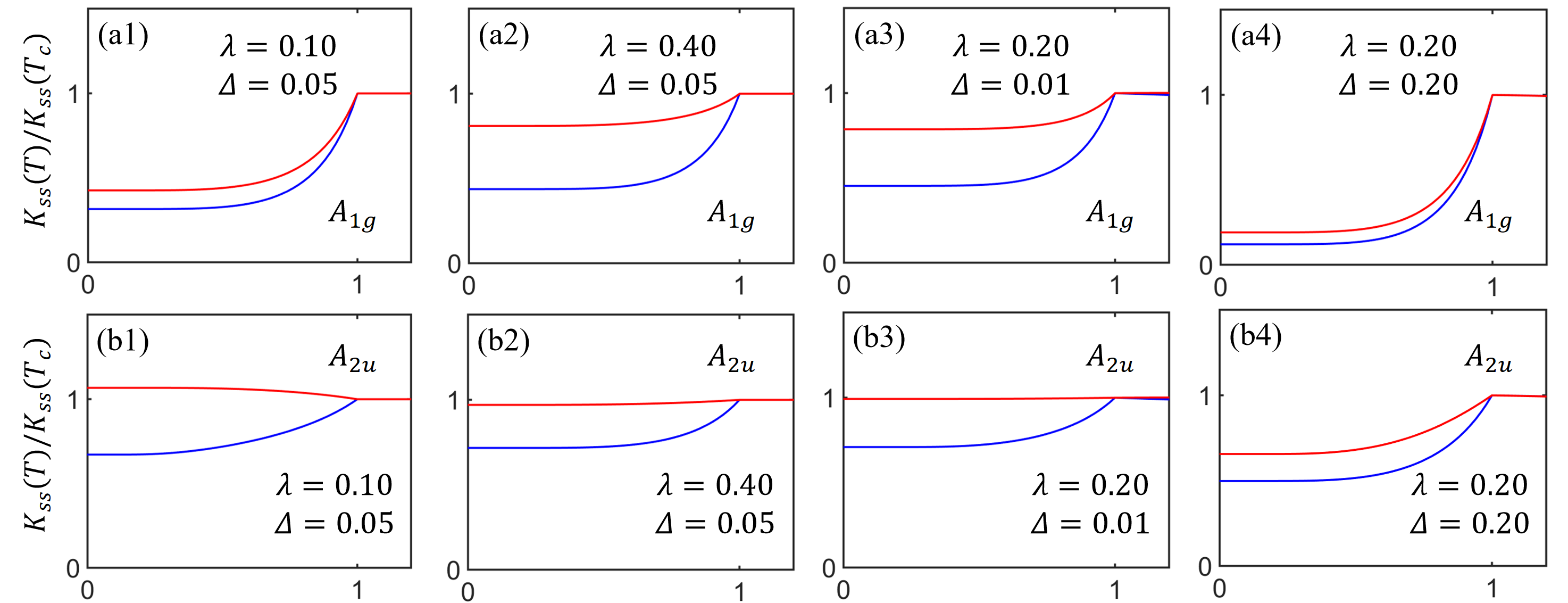}
	\caption{\label{SM_Knight_compare} (color online) The Knight shift for the $A_{1g}$ and $A_{2u}$ states in different conditions. In the calculations, the other parameters are chosen as $\{ t, t^\prime, \mu \} = \{ 1.0, 0.8, 0.3 \}$. The symbols in the figures are the same with that in the main text. }
\end{figure}

In the main text, we claim that for the system in our consideration, the Knight shift can be affected by the SOC  and the interband pairing. Here, we present more numerical results. We mainly consider the $A_{1g}$ and $A_{2u}$ states, since the $B_{2g}$ state can only appear in the weak SOC  condition (if the SOC  is strong, the $B_{2g}$ state is nodal and cannot be the ground state) where the Knight shift is qualitatively the same with that in the main text. As indicated in Fig.\ref{SM_Knight_compare} and the results shown in the main text, in the weak pairing condition, i.e. $\Delta << \lambda$, the Knight shift for the two states merely changes quantitatively as the SOC  varies. The $A_{2u}$ state can be distinguished from the other states through the unsuppressed $K_{zz}$. However, if the pairing is strong, i.e. $\Delta \sim \lambda$, the interband pairing changes the results qualitatively and it suppresses $K_{zz}$ in the superconducting state. Therefore, in the strong pairing condition, it is not a good choice to use the Knight shift to distinguish the different states.

\subsection{In-plane upper critical field}
For the superconducting ground states in the phase diagram in the main text, the in-plane upper critical field of spin-triplet $A_{2u}$ state is larger than the Pauli limit. Here, we show this by carrying out rough numerical simulations for the pairing orders in presence of the in-plane magnetic field. We consider the following Hamiltonian
\begin{eqnarray}\label{SM_Hamiltonian_field}
\mathcal{H}_{total} = \mathcal{H}_{eff} + \mathcal{H}_{int} + \mathcal{H}_{mag},
\end{eqnarray}
where $\mathcal{H}_{mag} = B s_1 \sigma_0$ is the in-plane Zeeman field, and $\mathcal{H}_{eff}$ and $\mathcal{H}_{int}$ are shown in the main text. In the calculations, we set the parameters as, $\{ t, t^\prime, \mu \lambda \} = \{ 1.0, 0.8, 0.3, 0.2 \}$ for the $A_{1g}$ and $A_{2u}$ states, and $\{ t, t^\prime, \mu \lambda \} = \{ 1.0, 0.8, 0.3, 0.1 \}$ for the $A_{1g}$ for the $B_{2g}$ state. For the Hamiltonian in Eq.\eqref{SM_Hamiltonian_field}, we solve the superconducting gap equation ${\bf \Delta({\bf k}) } = -\frac{1}{N} \sum_{{\bf k^\prime}} \mathcal{U} F({\bf k^\prime}, \tau=0)$ itinerantly. In solving the gap equation, for the $A_{1g}$ channel we choose $U = -0.8$ which leads to pairing order in the absence of the Zeeman field $\Delta_{0, A_{1g}} = 0.06$; and for the $B_{2g}$ and $A_{2u}$ channels, we set $V = -1.0$ corresponding to the pairing orders in the absence of the Zeeman field $\Delta_{0, B_{2g}} = 0.11$ and $\Delta_{0, A_{2u}} = 0.07$. Turning on the in-plane magnetic field, we get the pairing orders in Fig.\ref{SM_field}. Obviously, for the parameters chosen in the above, the pairing orders for the $A_{1g}$ and $B_{2g}$ states vanish at $B \sim \Delta_0$, while the $A_{2u}$ state has the upper critical field $B \sim 5\Delta_0$.

\begin{figure}[!htbp]
	\centering
	\includegraphics[width=0.4\linewidth]{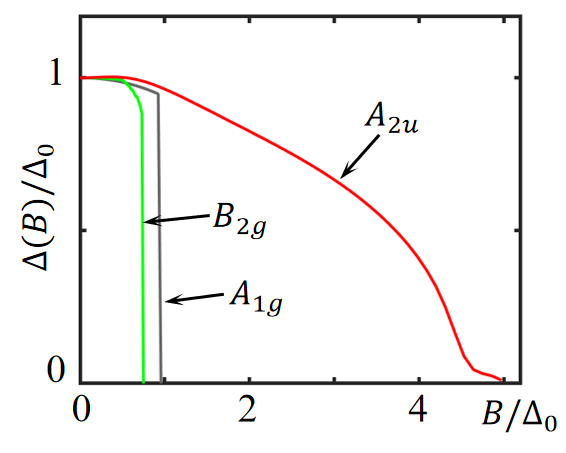}
	\caption{\label{SM_field} (color online) By solving the Hamiltonian in Eq.\eqref{SM_Hamiltonian_field}, we get the pairing orders in the presence of the in-plane magnetic field. }
\end{figure}

\end{widetext}


\begin{thebibliography}{reference}
\bibitem{footnote2}
The matrix form of the mirror symmetry and the nearly fully polarized spin near the Brillouin zone boundary, is closely related to the nonsymmorphic symmetry. To show this, in the SM we analyze the condition near the Brillouin zone center where the spin polarization turns out to be vanishing samll. This is because the nonsymmorphic symmetry has nontrivial effect merely on the Brillouin boundary, and it is simply equivalent to the point group symmetry at the Brillouin zone center.
\bibitem{footnote3}
Strictly speaking, the superconducting orders refer to leading order approximation of the momentum-dependent superconducting orders. More detailed analysis is shown in the SM.
\end{thebibliography}
\end{document}